\documentclass[11pt,reqno]{amsart}
\usepackage{amssymb,amsmath}
\usepackage{color} 
\usepackage{accents} 
\usepackage{texdraw}
\usepackage{eucal}
\usepackage{slashed} 
\usepackage{epic,epsfig}
\usepackage{graphicx}

\numberwithin{equation}{section}
\DeclareMathAccent{\wtilde}{\mathord}{largesymbols}{"65}
\DeclareMathAccent{\what}{\mathord}{largesymbols}{"62}

\def\m@th{\mathsurround=0pt}
\mathchardef\bracell="0365 
\def\upbrall{$\m@th\bracell$}
\def\undertilde#1{\mathop{\vtop{\ialign{##\crcr
    $\hfil\displaystyle{#1}\hfil$\crcr
     \noalign
     {\kern1.5pt\nointerlineskip}
     \upbrall\crcr\noalign{\kern1pt
   }}}}\limits}
   
\mathchardef\hatcell="0362
\def\dobrall{$\m@th\hatcell$}
\def\underhat#1{\mathop{\vtop{\ialign{##\crcr
    $\hfil\displaystyle{#1}\hfil$\crcr
     \noalign
     {\kern1.5pt\nointerlineskip}
     \dobrall\crcr\noalign{\kern1pt
   }}}}\limits}   
   
\def\theequation{\arabic{section}.\arabic{equation}}

\newcommand{\wh}{\widehat}
\newcommand{\wt}{\widetilde}
\newcommand{\ut}{\undertilde}
\newcommand{\uh}{\underhat}

\newcommand{\bblu}{\begin{color}{blue}}
\newcommand{\bred}{\begin{color}{red}}
\newcommand{\ecl}{\end{color}}

\newcommand{\bA}{\boldsymbol{A}} 
\newcommand{\bB}{\boldsymbol{B}} 
\newcommand{\bC}{\boldsymbol{C}} 
 
\newcommand{\bE}{\boldsymbol{E}} 
\newcommand{\bF}{\boldsymbol{F}} 
\newcommand{\bG}{\boldsymbol{G}} 
\newcommand{\bH}{\boldsymbol{H}}

\newcommand{\bK}{\boldsymbol{K}} 
\newcommand{\bL}{\boldsymbol{L}} 
\newcommand{\bM}{\boldsymbol{M}}

\newcommand{\bS}{\boldsymbol{S}} 
\newcommand{\bT}{\boldsymbol{T}}

\newcommand{\bZ}{\boldsymbol{Z}} 
\newcommand{\aR}{\alpha}
\newcommand{\bb}{\beta}
\newcommand{\gm}{\gamma}

\newcommand{\sg}{\sigma}

\newcommand{\kp}{\kappa}
\newcommand{\ld}{\lambda}

\newcommand{\oa}{\omega}
\newcommand{\be}{\begin{equation}}
\newcommand{\ee}{\end{equation}}
\newcommand{\bea}{\begin{eqnarray}}
\newcommand{\eea}{\end{eqnarray}}
\newcommand{\bse}{\begin{subequations}}
\newcommand{\ese}{\end{subequations}}
\newcommand{\nn}{\nonumber}

\newcommand{\ol}{\overline}
\newcommand{\ogm}{\overline{\gamma}}
\newcommand{\oS}{\overline{S}}
\newcommand{\obS}{\overline{\bS}}
\newcommand{\obG}{\overline{\bG}}

\newcommand{\bh}{\boldsymbol{h}}

\newcommand{\bxi}{{\boldsymbol \xi}}

 \newcommand{\ar}{\alpha} 
 \newcommand{\bgreen}{\begin{color}{green}}

\renewcommand{\theequation}{\thesection.\arabic{equation}}
\newcounter{alphabetical}
\newcounter{alphappendix}

\usepackage{pstricks}
\usepackage{calrsfs}
\usepackage{pifont}
\usepackage{amsbsy}
\usepackage{epic}

\newcommand{\bss}{\boldsymbol{s}}
\newcommand{\bchi}{{\boldsymbol \chi}}

\begin{document}
\title[ Discrete elliptic isomonodromic deformation problems]
{On elliptic Lax pairs and isomonodromic deformation systems for elliptic lattice equations}

\author{Frank Nijhoff and Neslihan Delice}
\address{
School of Mathematics\\
University of Leeds\\ 
Leeds LS2 9JT\\ 
United Kingdom \\ }
\email{F.W.Nijhoff@leeds.ac.uk  (corresponding author).}

\thanks{FWN is partially supported by EPSRC grant EP/I038683/1, while ND was supported by the Turkish Ministry of National Education.}

\begin{abstract}
In a previous article \cite{nesliFrankRin} a novel class of elliptic Lax pairs for integrable lattice equations was introduced. 
The present article proposes a de-autonomisation of those Lax pairs leading to a class of elliptic discrete isomonodromic deformation problems. 
We analyse the systems of compatibility conditions using some (possibly novel) higher order elliptic identities. 
\end{abstract}

\maketitle

\begin{flushright}
\emph{In honour of Professor Noumi for the occasion of his 60th birthday} 
\end{flushright}

\section{Introduction}
\setcounter{equation}{0} 

Ever since Richard Fuchs in 1905, \cite{Fuch2}, investigated the problem posed by his father, Lazarus Fuchs, namely to describe the iso-monodromic deformation of 
a linear second order ordinary differential equation (ODE) with 3 fixed and one moving regular singularity, and thus discovered for the first time the nonlinear differential equation 
for an apparent singularity that has to be included in the coefficients, which is now known by the name Painlev\'e VI equation (P$_{\rm VI}$), the study of linear 
systems of equations giving rise to nonlinear equations as compatibility conditions has formed a main theme in the subject called \emph{integrable systems}. 
Such systems of linear equations are nowadays usually loosely called \emph{Lax pairs}, referring to the work by Peter Lax, \cite{Lax}, on the general structure of 
the linear system found earlier by Gardner, Greene, Kruskal and Miura giving rise to the Korteweg-de Vries (KdV) equation. Richard Fuchs' problem constitutes 
probably the first ever example of a non-trivial Lax pair, although as an \emph{isomonodromic deformation problem} it is of a different type than the 
`autonomous' Lax pairs that have become prominent in the theory of soliton systems and integrable PDEs and their discrete counterparts, cf. e.g. \cite{AC}.  

The history of isomonodromic deformation theory, in connection with Painlev\'e type equations, has undergone some spectacular developments especially in recent years. 
R.Fuchs' 1905 paper\footnote{The discovery of P$_{\rm VI}$ in this paper was commemorated in the special issue \cite{CJMNN} where also some of the early history was summarised.} was 
followed up by his more often cited paper of 1907, \cite{Fuch}, as well as by the work  Ludwig Schlesinger,\cite{Schlesin}, on the isomonodromic deformations of more general matrix 
systems. 
Furthermore, Ren\'e Garnier in \cite{Garnier} extended Fuchs' scheme to higher order, allowing multiple moving singularities, leading to coupled systems of ODEs 
(one for each moving singularity), which are compatible through some additional partial differential relations. It is worth mentioning that the Garnier system amounts to what in 
modern terminology would be called a \textit{Painlev\'e VI hierarchy}. Fast forward almost seventy years, when the isomonodromic deformation theory in the sense of 
Painlev\'e equations, was rediscovered on the one hand in the seminal work of the Kyoto school, \cite{Jim1,Jim2,Jim3} and by Flaschka and Newell \cite{FlasNew}, and where the 
context of reductions (from integrable partial differential equations) allowed systematic constructions of isomonodromic deformation problems using the 
techniques of integrable systems. An overview of how isomonodromic deformation problems in combination with the Riemann-Hilbert techniques was used to access the transcendental solutions 
of Painelv\'e equations can be found in the monographs \cite{ItsNovok,FIKN}.  

In the early 1990s, when \emph{discrete Painlev\'e equations} made their appearance, discrete isomonodromic deformation problems were provided and studied in \cite{NP,FIK,PNGR,JBH}.  
The first isomonodromic deformation problem of $q$-difference type was provided in \cite{PNGR} for a $q$-version of the PIII equation. The latter, in fact, in its most 
general alternate form is equivalent to the $q$-version of PVI of Jimbo and Sakai, \cite{JimSakai}, who provided a $q$-isomonodromic deformation problem closer 
in spirit to the Fuchs' system for the continuous PVI equation, (cf. also \cite{Carg} for the connection between these two different isomonodromic deformation problems of 
$q$-difference type). The first proposals for the higher-order analogues, i.e., discrete Garnier type systems, were given in \cite{NW,TongasNij2}, while \cite{Sakai2} 
delivered a full $q$-analogue of the original Garnier system. The general theory of discrete isomonodromic deformation theory was developed in various 
recent papers, \cite{ArinBor,Bor,Krich}, where the Riemann-Hilbert aspects go back to work by Birkhoff and collaborators in the early 20th century, \cite{Birk,Birk2}.   
    
In this paper we present a general class of isomonodromic deformation problems which form (in some sense) the non-autonomous counterparts of the elliptic Lax systems studied in the article \cite{nesliFrankRin}. 
In the spirit of the early paper \cite{PNGR}, cf. also \cite{Carg}, where we introduced the {\it{ de-autonomization}} procedure for differential and $q$-difference Lax pairs which 
were obtained from periodic reductions of integrable lattice equations, leading to Lax pairs for associated discrete Painlev\'e equations, we will introduce isomonodromic  deformation 
problems of elliptic type. A general theory of elliptic type isomonodromic deformation problems, and the associated Riemann-Hilbert problem, was presented in \cite{Krich}, but 
the compatibility conditions were not worked out in that paper. On the other hand, the problem of finding a Lax pair for the famous elliptic discrete Painlev\'e equation of Sakai, 
\cite{Sakai}, was addressed by several authors in recent years, \cite{Yamada,Rains}, cf. also \cite{NoumTsujYam}, based on the birational geometry behind the Painlev\'e equations 
(see \cite{KajNoumYam} for a recent review). We mention, in passing, that in the continuous case, various elliptic isomonodromic deformation problems were studied 
in the literature \cite{{korsamt},{Takasaki}}, essentially going back to the pioneering work by Okamoto in the 1970s, \cite{{Okamoto2},{Okamoto3},{Okamoto4}}. 
We point out that in contrast to the above works on elliptic isomonodromic systems, our approach is very natural from the perspective of the lattice systems studied in 
\cite{nesliFrankRin} and is essentially based on the exploration of elliptic addition formulae. Along the way we present some seemingly novel elliptic identities of 
arbitrary order (i.e, in terms of arbitrary number of free arguments) which facilitate the analysis of the compatibility conditions. 
We will present the general scheme of equations, for arbitrary rank and order, and highlight some explicit cases for the sake of illustration.

\section{Autonomous elliptic Lax pairs for Integrable lattice systems}

In a previous article \cite{nesliFrankRin} a general elliptic Lax scheme of rank $N$, generalizing a novel Lax representation of Adler's lattice equation (Q4),\cite{Adler2}, was presented. 
In this section, mostly for the sake of introducing the notations, we outline the main construction of the corresponding autonomous lattice systems. For the basic definitions and standard formulae regarding the 
Weierstrass family of elliptic functions $\sg(x)$, $\zeta(x)$ and $\wp(x)$ we refer to Appendix A.  

We consider a compatible linear system for a $N$-component vector function $\bchi_\kp=\bchi_\kp(n,m)$, depending on a complex-valued spectral parameter $\kp$ (we 
prefer to indicate the dependence on the spectral parameter by means of an index rather than as a function argument to make the formulae more transparent), as well as on discrete variables $n,m$. In fact, the dependent variables of the systems under consideration, such as the functions $\xi_i(n,m)$ below, are functions of discrete independent variables $n$ and $m$, and for convenience we denote 
the unit shifts in these variables (i.e. elementary shifts along the lattice) by the notation:
\[ \wt{\xi}_i(n,m):=\xi_i(n+1,m)\  , \quad \wh{\xi}_i(n,m):=\xi_i(n,m+1)\  , \] 
while multiple shifts are denoted, e.g. by $\wh{\wt{\xi}}_i(n,m):=\xi(n+1,m+1)$, etc. 
The linear system (Lax pair) is given by the simultaneous pair of shift relations: 
\begin{equation}\label{eq:LatticeLax}  
\wt{\bchi}_\kp=\bL_\kp\,\bchi_\kp\  ,\quad \wh{\bchi}_\kp=\bM_\kp\,\bchi_\kp\  ,
\end{equation}
defining horizontal and vertical shits of the vector function $\bchi_\kp$, according to the diagram: \vspace{.1cm}

\begin{center}

\begin{figure}[ht]
\setlength{\unitlength}{0.00029745in}
\begingroup\makeatletter\ifx\SetFigFont\undefined%
\gdef\SetFigFont#1#2#3#4#5{%
  \reset@font\fontsize{#1}{#2pt}%
  \fontfamily{#3}\fontseries{#4}\fontshape{#5}%
  \selectfont}%
\fi\endgroup%
{\renewcommand{\dashlinestretch}{30}
\begin{picture}(4182,3705)(0,-10)
\put(652,3218){\circle*{144}}
\put(3420,3195){\circle*{144}}
\put(675,495){\circle*{144}}
\put(3446,508){\circle*{144}}
\thicklines
\drawline(675,3195)(3240,3195)
\drawline(675,3195)(3240,3195)
\drawline(3000.000,3135.000)(3240.000,3195.000)(3000.000,3255.000)(3000.000,3135.000)
\drawline(3420,3150)(3420,675)
\drawline(3420,3150)(3420,675)
\drawline(3360.000,915.000)(3420.000,675.000)(3480.000,915.000)(3360.000,915.000)
\dashline{90.000}(630,3105)(630,675)
\drawline(570.000,915.000)(630.000,675.000)(690.000,915.000)(570.000,915.000)
\dashline{90.000}(675,495)(3330,495)
\drawline(3090.000,435.000)(3330.000,495.000)(3090.000,555.000)(3090.000,435.000)
\put(495,3510){\makebox(0,0)[lb]{$\bchi$}}
\put(3420,3465){\makebox(0,0)[lb]{$\wt{\bchi}$}}
\put(300,180){\makebox(0,0)[lb]{$\wh{\bchi}$}}
\put(3690,270){\makebox(0,0)[lb]{$\wh{\wt{\bchi}}$}}
\put(1665,3465){\makebox(0,0)[lb]{$\bL$}}
\put(3690,1980){\makebox(0,0)[lb]{$\wt{\bM}$}}
\put(0,1845){\makebox(0,0)[lb]{$\bM$}}
\put(1800,0){\makebox(0,0)[lb]{$\wh{\bL}$}}
\end{picture}
\caption{Lattice compatibility configuration.}\label{lattcompatconfig}
}
\end{figure}
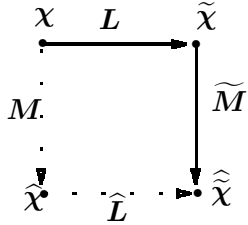

\end{center}
The compatibility condition (discrete zero-curvature condition) leads to the matrix relation 
\begin{equation}\label{eq:ZCcond} 
\wh{\bL}_\kp\bM_\kp=\wt{\bM}_\kp \bL_\kp\  . 
\end{equation}
\vspace{.2cm} 
The Lax matrices\footnote{We prefer to indicate the dependence of these matrices on the spectral parameter $\kp$ by 
an index notation, even though $\kp$ is generally a complex valued quantity (being a uniformising variable for the 
Weierstrass elliptic curve), in order to make the $\kp$ dependence clearly visible.} $\bL_\kp$ and $\bM_\kp$ attached to the edges are assumed to be of the form: 
\bse\label{eq:LM}\begin{eqnarray}
&& (\bL_\kp)_{i,j} = \Phi_{\kp}(\wt{\xi}_i-\xi_j-\aR) h_j\   , \label{eq:L} \\
&& (\bM_\kp)_{i,j} = \Phi_{\kp}(\wh{\xi}_i-\xi_j-\bb) k_j\   , \quad (j=1,\dots,N) \label{eq:M} 
\end{eqnarray}\ese 
where the coefficients $h_j$ and $k_j$ are assumed to be independent of the spectral parameter $\kp$, as are the functions 
$\xi_i=\xi_i(n,m)$ appearing in the argument of the function $\Phi_\kp$ which is defined as  
\begin{equation}\label{eq:Phi} 
 \Phi_\kp(x):=\frac{\sg(\kp+x)}{\sg(\kp)\,\sg(x)}\  . 
 \end{equation}  
 The parameters $\alpha$ and $\beta$ in \eqref{eq:LM} are constants, but as they are associated with the lattice directions we refer 
 to them as lattice parameters. 
   
Working out the compatibility condition \eqref{eq:ZCcond} by means of the basic addition formulae given in Appendix A, in particular \eqref{eq:12}, the consistency requirement gives rise to
\begin{eqnarray*}
&&  \sum_{l=1}^N \,\wh{h}_l k_j \left[\zeta(\wh{\wt{\xi}}_i-\wh{\xi}_l-\aR)+\zeta(\wh{\xi}_l-\xi_j-\bb)
+ \zeta(\kp)-\zeta(\kp+\wh{\wt{\xi}}_i-\xi_j-\aR-\bb)\right] = \nn \\
&& =\quad  \sum_{l=1}^N \,\wt{k}_l h_j \left[\zeta(\wh{\wt{\xi}}_i-\wt{\xi}_l-\bb)+\zeta(\wt{\xi}_l-\xi_j-\aR)
+ \zeta(\kp)-\zeta(\kp+\wh{\wt{\xi}}_i-\xi_j-\aR-\bb)\right] \nn  \\
&& \qquad \qquad\qquad  (i,j=1,\dots,N)\  .
\end{eqnarray*} 
Due to the (arbitrary) dependence on the spectral parameter $\kp$ these equations separate into two parts:
\bse\begin{eqnarray}
&& \left( \sum_{l=1}^N \wh{h}_l\right)k_j= \left(\sum_{l=1}^N \wt{k}_l \right) h_j\quad,\quad
(j=1,\dots,N)\  , \\
&&  \sum_{l=1}^N \,\wh{h}_l\left[\zeta(\wh{\wt{\xi}}_i-\wh{\xi}_l-\aR)+\zeta(\wh{\xi}_l-\xi_j-\bb) \right]k_j
= \sum_{l=1}^N \,\wt{k}_l\left[\zeta(\wh{\wt{\xi}}_i-\wt{\xi}_l-\bb)+\zeta(\wt{\xi}_l-\xi_j-\aR)\right]h_j \nn \\
&& \qquad\qquad\qquad (i,j=1,\dots,N)\  .
\end{eqnarray}\ese 
From this system of equations we want to extract a closed form system of lattice equations for 
the main dependent variables $\xi_i(n,m)$, eliminating the coefficient variables $h_i$, $k_i$.  
In order to do so we have to distinguish now between two cases which we referred to in \cite{nesliFrankRin} as 
\textit{Landau-Lifschitz type} (spin non-zero) and \textit{Krichever-Novikov type} (spin zero) respectively, 
depending on whether $\sum_l h_l$ non-vanishing or zero. The latter case (for $N=2$) corresponds to the 
case of Q4, in which case we obtain readily the 3-leg form of that equation. The main part of the 
paper \cite{nesliFrankRin} was dedicated to analysing the higher-rank situation of that case, 
deriving a coupled set of implicit lattice equations, analogous to the 3-leg form, for $N=3$, we 
will not summarize those results here as they require a lot of additional notation.

In the case where $\sum_l h_l\neq 0$ we have that the variables $h_j$, $k_j$ are proportional to each other, ~$k_j=\rho h_j$~, 
and after summation we obtain the (multiplicative) conservation law:
\begin{equation}\label{eq:LLconserv} 
\frac{\sum_{l=1}^N \wh{h}_l}{\sum_{l=1}^N h_l}=\frac{\sum_{l=1}^N \wt{k}_l}{\sum_{l=1}^N k_l}\  , 
\end{equation} 
so that the Lax equations reduce to 
\[
\sum_{l=1}^N \,\left[ \zeta(\wh{\wt{\xi}}_i-\wh{\xi}_l-\ar) \rho\wh{h}_l-\zeta(\wh{\wt{\xi}}_l-\wt{\xi}_j-\bb) \wt{k}_l \right] =
\sum_{l=1}^N \,\left[ \zeta(\xi_j-\wh{\xi}_l+\bb) \rho\wh{h}_l -\zeta(\xi_j-\wt{\xi}_l+\ar) \wt{k}_l \right]\  .
\]
($i,j=1,\dots,N$). 

Under the further assumption that the Centre of Mass (CoM) motion obeys the equation   
\begin{equation}\label{eq:CMmotion} 
\wt{\Xi}+\wh{\Xi}=\wh{\wt{\Xi}}+\Xi\ , \quad {\rm where} \quad
\Xi:=\sum_{j=1}^N \xi_j\,
\end{equation} 
we can analyse the Lax equations most conveniently by considering the following elliptic function and its 
expansion in terms of Weierstrass $\zeta$-functions, using the elliptic Lagrange interpolation formula \eqref{eq:Lagr2} in Appendix B: 
\begin{eqnarray*}
&& F(\xi):= \prod_{l=1}^N\,\frac{\sg(\xi-\wh{\wt{\xi}}_l)\sg(\xi-\xi_l-\ar-\bb)}{\sg(\xi-\wh{\xi}_l-\ar)\sg(\xi-\wt{\xi}_l-\bb)}  \\
&& = \sum_{l=1}^N \left[ \zeta(\xi-\wh{\xi}_l-\ar)- \zeta(\begin{color}{blue}\eta\end{color}-\wh{\xi}_l-\ar)\right] H_l
 + \sum_{l=1}^N \left[ \zeta(\xi-\wt{\xi}_l-\bb)- \zeta(\begin{color}{blue}\eta\end{color}-\wt{\xi}_l-\bb)\right] K_l\  . 
\end{eqnarray*}
The latter holds as an identity for any four sets of variables $\xi_l$, $\wh{\xi}_l$, $\wt{\xi}_l$,
$\wh{\wt{\xi}}_l$ such that the relation \eqref{eq:CMmotion} for their sums holds. Here $\eta$ denotes 
any one of the zeroes (i.e., $\wh{\wt{\xi}}_i$ or $\xi_i+\ar+\bb$), and  we have the explicit expressions for the coefficients: 
\bse\label{eq:LLcoeffs}\begin{eqnarray}
H_l&=& \frac{\prod_{k=1}^N \sg(\wh{\xi}_l-\wh{\wt{\xi}}_k+\ar)\sg(\wh{\xi}_l-\xi_k-\bb)}
{\left[\prod_{k=1}^N \sg(\wh{\xi}_l-\wt{\xi}_k+\ar-\bb)\right]\prod_{k\neq l} \sg(\wh{\xi}_l-\wh{\xi}_k)}\  ,  \\
K_l&=& \frac{\prod_{k=1}^N \sg(\wt{\xi}_l-\wh{\wt{\xi}}_k+\bb)\sg(\wt{\xi}_l-\xi_k-\ar)}
{\left[\prod_{k=1}^N \sg(\wt{\xi}_l-\wh{\xi}_k+\bb-\ar)\right]\prod_{k\neq l} \sg(\wt{\xi}_l-\wt{\xi}_k)} \  .
\end{eqnarray}\ese 
We note that by construction the coefficients obey the identity
$ \sum_{l=1}^N(H_l+K_l)=0\  . $

Using the identities above, taking $\xi=\wh{\wt{\xi}}_i$, $\eta=\xi_j+\ar+\bb$ in $F(\xi)$, and comparing the 
result with the Lax equations, we can identify:
\[
t H_l=\rho\wh{h}_l\quad ,\quad -tK_l=\wt{k}_l=\wt{\rho}\wt{h}_l\quad ,\quad l=1,\dots, N\  ,
\]
with $t$ an arbitrary proportionality factor.  Thus, inserting the explicit expressions
for $H_l$ and $K_l$ we obtain a system of $N+2$ equations for the $N+2$ unknowns: $\xi_1,\dots,\xi_N$,
$\rho$, $t$. 
This comprises the set of equations
\[
\frac{\wt{t}}{\wt{\rho}}\wt{H}_l+\frac{\wh{t}}{\wh{\wt{\rho}}} \wh{K}_l=0\ ,\quad (l=1,\dots,N)\  ,  
\quad \wt{\Xi}+\wh{\Xi}=\wh{\wt{\Xi}}+\Xi\ ,
\] 
which yields the system of $N$ 7-point equations:  
\begin{equation}\label{eq:7pteqs}
\prod_{k=1}^N\,\frac{\sg(\xi_l-\wt{\xi}_k+\alpha)\,\sg(\xi_l-{\uh{\xi}}\!\!\!\phantom{a}_k-\beta)\,
\sg(\xi_l-{\ut{\wh{\xi}}}\!\!\!\phantom{a}_k+\gamma)}
{\sg(\xi_l-\wh{\xi}_k+\beta)\,\sg(\xi_l-{\ut{\xi}}\!\!\!\phantom{a}_k-\alpha)\,\sg(\xi_l-{\uh{\wt{\xi}}}\!\!\!\phantom{a}_k-\gamma)}=p
\end{equation}
for $N+1$ variables $\xi_i$ ($i=1,\dots,N$) and $p=-\ut{t}\,\uh{\rho}/(\uh{t}\,\rho)$, supplemented
with the relation \eqref{eq:CMmotion}, which fixes the CoM dynamics. The under-accents $\ut{\cdot}$ and 
$\uh{\cdot}$ in \eqref{eq:7pteqs} denote reverse lattice shifts: ${\ut{\xi}}\!\!\!\phantom{a}_i(n,m)=\xi_i(n-1,m)$, ${\uh{\xi}}\!\!\!\phantom{a}_i(n,m)=\xi_i(n,m-1)$. We note that the implicit system of P$\Delta$Es arises from 
the following Lagrangian: 
\[ \mathcal {L}= \sum_{i,j}\,\left[ f(\xi_i-\wt{\xi}_j+\alpha)-f(\xi_i-\wh{\xi}_j+\beta)-
f(\wh{\xi}_i-\wt{\xi}_j+\alpha-\beta) \right] - \ln|p|\,\Xi \]    
in which the function $f$ is the elliptic dilogarithm ~$f(x)=\int^x\,\ln\,\sg(\xi)\,d\xi$~. For reasons that we will
not go into here we expect the system of equations for $N=2$ to constitute an implicit form of a lattice 
version of the Landau-Lifschitz (LL) equations, in a similar way as the 3-leg equations are an implicit form 
of Q4. Lattice versions of the LL equations were proposed in \cite{NijPap,AdYam,Adler3}, but so far no relation 
between these different proposals has been established, nor is anything known yet about the solution structure of those models. 

\paragraph{\bf Remark:} The one-step periodic reduction,  ~$\wt{\chi}_\kappa=\lambda\chi_\kappa$~, reduces the 
Lax system \eqref{eq:LatticeLax} to the Lax pair of the discrete-time Ruijsenaars system that was 
constructed in \cite{NRK}. In that case the Lax equation ~$\wh{\bL}_\kappa\bM_\kappa=\bM_\kappa\bL_\kappa$ ~ 
yields an implicit system of second order O$\Delta$Es which constitutes the discrete-time equations of 
motion for the discrete-time Ruijsenaars system. In the simplest case, these will re-emerge as the autonomous 
limit of the systems we will consider next in the simplest case. In a sense the results described in the remainder 
of the paper could be thought of as constituting a similarity reduction of the system of 7-point equations \eqref{eq:7pteqs}. 
We will come back to this in the Conclusions.

\section{Elliptic isomonodromic deformation scheme}
\label{sec:General}

In this section, we introduce a general class of isomonodromic deformation problems on the torus, inspired by the form of the 
lattice systems of the previous section. 

Using a similar notation as before we now consider the compatibility of the system of linear equations
\bse\label{eq:Laxmat}\begin{eqnarray}
\bchi_{\kp+\tau} &=&\bT_\kp\,\bchi_\kp\ ,\label{eq:Laxmatb} \\ 
\wt{\bchi}_\kp &=& \bL_\kp\,\bchi_\kp\  ,\label{eq:Laxmata}
\end{eqnarray}\ese
where the first relation \eqref{eq:Laxmatb} is a linear first order difference equation on the torus, i.e. defining a shift over a fixed increment 
$\tau$ in the uniformizing spectral variable $\kp$, while the second relation \eqref{eq:Laxmata} defines a shift $\wt{\phantom{a}}$ of the 
vector function $\bchi_\kp$ in some additional discrete variable $n$ as according to \eqref{eq:LatticeLax}. Eventually we will equip the 
system \eqref{eq:Laxmat} with additional lattice directions, in additional variables such as $m$ as in section 2, each of which corresponding to 
a linear equation such as \eqref{eq:Laxmata}. 

\subsection{General scheme}

We take the matrices $\bL_\kp$ and $\bT_\kp$ of the form
\bse \label{eq:Laxxmatricesss} \begin{eqnarray} 
 (\bL_\kp)_{i,j} &=& H_{i,j}\ \sg(\kp) \,\Phi_{\kp}(\wt{\xi}_i-\xi_j-\alpha)\   ,\label{eq:Laxxmatricesssa}  \\
 (\bT_{\kp})_{i,j} &=& \sum_{l_1,\dots,l_{m-1}}\,S_{i,j}^{(l_1,\dots,l_{m-1})}\,\prod_{\nu=1}^m 
 \sg(\kp-\kp_\nu)\,\Phi_{\kp-\kp_\nu}(\xi^{(\nu-1)}_{l_{\nu-1}}-\xi^{(\nu)}_{l_\nu}-\gm_\nu)\   , \label{eq:Laxxmatricesssb} \\
\phantom{a}  &\phantom{a} & \qquad \qquad  (i,j=1,\dots,N)  \nn
\end{eqnarray}\ese 
where the matrices $\bH=(H_{i,j})$ and the quantities $S_{i,j}^{l,l_1,\dots,l_{m-1}}$ remain to be specified, and in which we identify:
\[  \xi^{(0)}_{l_0}=\xi_i\ , \quad \xi^{(m)}_{l_m}=\xi_j\  . \] 
The parameters $\kp_\nu$, ($\nu=1,\dots,m$), are fixed, while the quantities $\gm_\nu$, ($\nu=1,\dots,m$), are assumed to be functions of the 
discrete variable $n$, the precise dependence of which follows from the analysis below. All quantities are assumed to be independent of $\kp$ unless 
explicitly indicated.  

The compatibility condition
\begin{equation}\label{eq:compat}
\wt{\bT}_\kp\,\bL_\kp= \bL_{\kp+\tau}\,\bT_\kp \  , 
\end{equation} 
Gives rise to 
\begin{eqnarray*}
&& \sum_{l_1,\dots,l_{m}}\,\wt{S}_{i,l_m}^{(l_1,\dots,l_{m-1})}\,H_{l_m,j}\,\left[\prod_{\nu=1}^m 
 \sg(\kp-\kp_\nu)\,\Phi_{\kp-\kp_\nu}(\wt{\xi}^{(\nu-1)}_{l_{\nu-1}}-\wt{\xi}^{(\nu)}_{l_\nu}-\wt{\gm}_\nu)\right]\,\sg(\kp)
 \Phi_{\kp}(\wt{\xi}_{l_m}-\xi_j-\alpha) = \\  
 &&= \sum_{l,l'_1,\dots,l'_{m-1}}\,H_{i,l}\,S_{l,j}^{(l'_1,\dots,l'_{m-1})} \sg(\kp+\tau)\Phi_{\kp+\tau}(\wt{\xi}_i-\xi_l-\alpha) 
 \prod_{\nu=1}^m 
 \sg(\kp-\kp_\nu)\,\Phi_{\kp-\kp_\nu}(\xi^{(\nu-1)}_{l'_{\nu-1}}-\xi^{(\nu)}_{l'_\nu}-\gm_\nu)\  \, 
\end{eqnarray*}
in which we set in addition to the above identifications: 
\[ \xi^{(0)}_{l'_0}=\xi_l\  , \quad \xi^{(m)}_{l'_{m}}=\xi_j\  . \] 
Using the relation $\Phi_\kp(\tau)\Phi_{\kp+\tau}(x)=\Phi_\kp(\tau+x)\Phi_\tau(x)$ on the right hand side, as well as the identity 
\eqref{eq:GenEllProd} we can rewrite both sides of the latter equality to yield: 
\begin{eqnarray*}
&& \sum_{l_1,\dots,l_{m}}\,\wt{S}_{i,l_m}^{(l_1,\dots,l_{m-1})}\,H_{l_m,j}\,\sum_{\nu'=1}^{m+1} 
\Phi_{\kp-\kp_{\nu'}}(\wt{\xi}_i-\xi_j-\ar-\wt{\gm})\prod_{\nu=1\atop \nu\neq\nu'}^{m+1} 
\,\Phi_{\kp_{\nu'}-\kp_\nu}(\wt{\xi}^{(\nu-1)}_{l_{\nu-1}}-\wt{\xi}^{(\nu)}_{l_\nu}-\wt{\gm}_\nu) = \\  
&&= \sum_{l,l'_1,\dots,l'_{m-1}}\,H_{i,l}\,S_{l,j}^{(l'_1,\dots,l'_{m-1})} \sg(\tau)\Phi_{\tau}(\wt{\xi}_i-\xi_l-\alpha) \\ 
&& \qquad\qquad \times \sum_{\nu'=0}^m
\Phi_{\kp-\kp_{\nu'}}(\wt{\xi}_i-\xi_j-\ar+\tau-\gm) \prod_{\nu=0\atop \nu\neq\nu'}^m 
\,\Phi_{\kp_{\nu'}-\kp_\nu}(\xi^{(\nu-1)}_{l'_{\nu-1}}-\xi^{(\nu)}_{l'_\nu}-\gm_\nu) \  , 
\end{eqnarray*}
where (in order to avoid having to separate the sums and products) we have introduced the notations:  
\[ \kp_0:=0\  , \quad \kp_{m+1}:=0\  , \quad \gm_{m+1}:=0\  , \quad \xi^{(-1)}_{l'_{-1}}=\wt{\xi}_i\  , 
\quad  \wt{\xi}^{(m+1)}_{l_{m+1}}=\xi_j+\alpha\  , \quad \gm:=\sum_{\nu=1}^m \gm_\nu\  .   \]
Setting now ~$\wt{\gm}=\gm-\tau$~ (implying that at this point we take the $\gm_\nu$ to depend on $n$ such that their sum is a linear 
function of the discrete variable) the terms depending on $\kp$ can be identified leading to the system of relations: 
\bse\label{eq:Ellsyst}\begin{eqnarray} 
&& \sum_{l_1,\dots,l_{m}}\,\wt{S}_{i,l_m}^{(l_1,\dots,l_{m-1})}\,H_{l_m,j}\,\prod_{\nu=1\atop \nu\neq\nu'}^{m+1} 
\,\Phi_{\kp_{\nu'}-\kp_\nu}(\wt{\xi}^{(\nu-1)}_{l_{\nu-1}}-\wt{\xi}^{(\nu)}_{l_\nu}-\wt{\gm}_\nu) = \nn \\  
&&= \sum_{l,l'_1,\dots,l'_{m-1}}\,H_{i,l}\,S_{l,j}^{(l'_1,\dots,l'_{m-1})} \sg(\tau)\Phi_{\tau}(\wt{\xi}_i-\xi_l-\alpha)\,\prod_{\nu=0\atop \nu\neq\nu'}^m 
\,\Phi_{\kp_{\nu'}-\kp_\nu}(\xi^{(\nu-1)}_{l'_{\nu-1}}-\xi^{(\nu)}_{l'_\nu}-\gm_\nu) \  , \nn \\ 
&& \qquad\qquad \qquad \nu'=1,\dots,m\ , \quad i,j=1,\dots, N, \label{eq:Ellsysta}
\end{eqnarray}
together with 
\begin{eqnarray} 
&& \sum_{l,l_1,\dots,l_{m-1}}\,\wt{S}_{i,l}^{(l_1,\dots,l_{m-1})}\,H_{l,j}\,\prod_{\nu=1}^{m} 
\,\Phi_{-\kp_\nu}(\wt{\xi}^{(\nu-1)}_{l_{\nu-1}}-\wt{\xi}^{(\nu)}_{l_\nu}-\wt{\gm}_\nu) = \nn \\  
&&= \sum_{l,l'_1,\dots,l'_{m-1}}\,H_{i,l}\,S_{l,j}^{(l'_1,\dots,l'_{m-1})} \sg(\tau)\Phi_{\tau}(\wt{\xi}_i-\xi_l-\alpha)\,\prod_{\nu=1}^m 
\,\Phi_{-\kp_\nu}(\xi^{(\nu-1)}_{l'_{\nu-1}}-\xi^{(\nu)}_{l'_\nu}-\gm_\nu) \  . \nn \\ 
&& \qquad\qquad \qquad  i,j=1,\dots, N, \label{eq:Ellsystb}
\end{eqnarray}\ese 
The fundamental system of relations \eqref{eq:Ellsyst} is the basis of further analysis. Like in the autonomous system described in the section 2, the coefficient 
matrices $H_{i,j}$ and, in this case, $S_{i,j}^{(l'_1,\dots,l'_{m-1})}$ have to be eliminated. To remain in the spirit of the previous case we will make some simplifying assumptions, 
for instance that the matrix of coefficient $(H_{ij})$ is of rank 1.  Eliminating the those coefficients would yield a system of of equations for the main quantities 
$\xi_i$ including the quantities $\xi_i^{(\nu)}$. For consistency also some `global' conditions may be needed on the latter quantities, such as certain restrictions on the sums 
$\Xi^{(\nu)}=\sum_l \xi_l^{(\nu)}$.

\subsection{Example: First order scheme}

In order to make the structure of the Lax system \eqref{eq:Ellsyst} more transparant, we will first illustrate them by means of some simpler examples, 
namely the cases where $m=1$ and $m=2$ in  \eqref{eq:Laxxmatricesssb}. 

\paragraph{\it First order scheme ($m=1$)} 
In this case we have the elliptic discrete isomonodromic system \eqref{eq:Laxmat} with Lax matrices of the form:  
\bse\label{eq:1storderLax}\begin{eqnarray}  
 (\bL_\kp)_{i,j}&=& H_{i,j}\ \sg(\kp) \,\Phi_{\kp}(\wt{\xi}_i-\xi_j-\alpha)\   ,  \label{eq:1storderLaxa} \\ 
 (\bT_{\kp})_{i,j}&=& S_{i,j}\ \sg(\kp-\kp_1)\,\Phi_{\kp-\kp_1}(\xi_i-\xi_j-\gamma)\ ,   \label{eq:1storderLaxb}
\qquad  (i,j=1,\dots,N)\ .   
\end{eqnarray}\ese  
The coefficients $H_{i,j}$, $S_{i,j}$ do not depend on the spectral parameter $\kappa$ and remain to be determined, while the $\xi_i=\xi_i(n)$ are 
the main independent variables.
We observe in this case that the forms of the matrices \eqref{eq:1storderLax} are reminiscent of those of the discrete zero-curvature Lax pair 
\eqref{eq:LM}, except that we include here extra factors $\sg(\kp)$ which turn out to be necessary in order to separate out the $\kp$-dependence 
in the consistency conditions\footnote{In the autonomous case those extra factors can be readily removed by means of a simple gauge transformation.}. 
Furthermore, the coefficients $h_j$, $k_j$ in \eqref{eq:LM} correspond to a rank 1 restriction on the coefficient matrices $\bH=(H_{ij})$ and 
$\bS=(S_{ij})$, which here we don't want to impose from the start in order to allow for more freedom in the analysis. We observe, furthermore, 
that in this case we have a single variable $\gamma$, which depends linearly on $n$ via the relation 
$$\wt{\gm}=\gm-\tau\quad \Rightarrow \quad \gm=\gm(n)=\gm(0)-n\tau\  , $$ 
on the discrete variables $n$. In this first order case the system \eqref{eq:Ellsyst} adopts the form:
\bse\label{eq:1stordsyst}\bea
&& \sum_{l=1}^N \wt{S}_{il} H_{lj}\,\Phi_{\kp_1}(\wt{\xi}_l-\xi_j-\alpha) 
=\sum_{l=1}^N H_{il}S_{lj}\,\sg(\tau)\Phi_{\tau}(\wt{\xi}_i-\xi_l-\alpha)\,\Phi_{\kp_1}(\tau+\wt{\xi}_i-\xi_l-\alpha)\ , \nn \\ 
&& \label{eq:1stordsysta}\\
&& \sum_{l=1}^N \wt{S}_{il} H_{lj}\,\Phi_{-\kp_1}(\wt{\xi}_i-\wt{\xi_l}-\wt{\gamma})=
\sum_{l=1}^N H_{il}S_{lj}\,\sg(\tau)\Phi_{\tau}(\wt{\xi}_i-\xi_l-\alpha)\,\Phi_{-\kp_1}(\xi_l-\xi_j-\gamma), \nn \\ 
&& \qquad\qquad\qquad i,j=1,\dots,N\ . \label{eq:1stordsystb}
\eea\ese

Let us first note that for $N=1$ (scalar case) this system of equations becomes quite simple and reduces to the equality
\[ \frac{\wt{S}}{S}=\frac{\sg(\tau)\,\Phi_\tau(\wt{\xi}-\xi-\alpha)\,\Phi_{\kp_1}(\tau+\wt{\xi}-\xi-\alpha)}{\Phi_{\kp_1}(\wt{\xi}-\xi-\alpha)}
= \frac{\sg(\tau)\,\Phi_\tau(\wt{\xi}-\xi-\alpha)\,\Phi_{-\kp_1}(-\gm)}{\Phi_{-\kp_1}(-\wt{\gm})}\  , \]
where $S=S_{11}$ and $\xi=\xi_1$. Multiplying out the denominators from the second equality, the latter reduces further by using the addition formula \eqref{eq:12} and yields the simple relation 
\[\zeta(\tau+\wt{\xi}-\xi-\alpha)+\zeta(\wt{\gm})-\zeta(\wt{\xi}-\xi-\alpha) -\zeta(\gm)=0\  , \] 
from which the parameter $\kp_1$ has disappeared. This last equation can be resolved by using \eqref{eq:zs} and yields the following two branches of solutions:  
\[ \wt{\xi}-\xi-\alpha+\gm\doteq 0 \qquad {\rm and}\qquad \wt{\xi}-\xi-\alpha-\wt{\gm}\doteq 0\  , \] 
(in which ``$\doteq 0$'' indicates that the equality holds modulo the period lattice of the Weierstrass elliptic functions). Thus, 
in this simple case we find that the dependent variable $\xi$ depends quadratically on the discrete variable $n$:
\begin{equation}\label{eq:N=1sol}
\xi(n)=\xi(0)+(\alpha\mp \gm(0)+\Omega)n \pm \tfrac{1}{2}n(n\mp 1)\tau\  , 
\end{equation} 
(in which $\Omega$ denotes any integer combination of the periods of the Weierstrass functions\footnote{In principle the choice of period $\Omega$ 
does not need to be fixed, but could alter under application of the map. We will not consider that possibility in this paper, but only note that 
the presence of the freedom of choosing periods in \eqref{eq:N=1sol} will not alter the dependence on $n$ in quantities as $u:=\wp(\xi(n)$ which would obey rational 
counterparts of the equations considered.}). 
This ``scalar Lax'' case is not quite trivial, even though it can be integrated explicitly, recalling that the representation involves 
functions $\xi(n)$ which appear in the arguments of elliptic functions, and that the quadratic dependence on the independent variable is 
reminiscent of certain cases of Painlev\'e type equations which can be linearized.  
 
To analyse the cases $N\geq 2$ it is convenient to introduce a somewhat unconventional matrix notation. For any two 
$N\times N$ matrices $\bA=(A_{ij})$ and $\bB=(B_{ij})$, let us introduce the operation of ``gluing'' two 
matrices,i.e., the entry-by-entry multiplication creating the ``glued'' matrix denoted by $[\bA\bB]$ 
having entries: 
\[ \left([\bA\bB]\right)_{ij}:= A_{ij}B_{ij}\  .    \]   
This allows us to rewrite \eqref{eq:1stordsyst} in the following short-hand way: 
\bse\label{eq:Nmatsyst}\begin{eqnarray}
&& \wt{\bS}\cdot[\bA_{\kp_1}\bH] = [\bA_{\kp_1+\tau}\bH]\cdot\bS\ , \label{eq:Nmatsysta}  \\
&& [\wt{\bG}_{-\kp_1}\wt{\bS}]\cdot\bH = 
[\bA_\tau\bH]\cdot[\bG_{-\kp_1}\bS]\ , \label{eq:Nmatsystb}
\end{eqnarray} \ese
where we have introduced the matrices  
\begin{eqnarray*}
(\bA_\kp)_{ij}&:=&\sigma(\kp)\Phi_{\kp}(\wt{\xi}_i-\xi_j-\alpha)\  , \nn\\
(\bG_\kp)_{ij}&:=&\sigma(\kp)\Phi_{\kp}(\xi_i-\xi_j-\gamma)\  . 
\end{eqnarray*} 
As in the autonomous case of section 2, we are interested primarily in the case that the matrix $\bH$ 
is of rank 1, in which case from \eqref{eq:Nmatsystb} we have that either the glued matrix $[\bA_\tau\bH]$ 
must be singular, or the matrix $[\bG_{-\kp_1}\bS]$ is singular. In the former case, as a consequence of the 
Frobenius determinant formula \eqref{eq:cauchy} of Appendix B, we must have that 
\[ \det(\bA_\tau)=0\quad \Rightarrow\quad \tau+\wt{\Xi}-\Xi-N\alpha=0\  , \quad {\rm with}\quad \Xi=\sum_{j=1}^N \xi_j\  . \]
For $N=2$ this condition on the sums of the $\xi$-variables is a sufficient condition for $[\bA_\tau\bH]$ to 
be of rank 1, as the glued matrix is, up to multiplication by diagonal matrices, a Frobenius 
(i.e., elliptic Cauchy) matrix. Since, as a consequence, in that case in general position the matrices 
$\bA_{\kp_1}$ and $\bA_{\kp_1+\tau}$ are invertible (provided we avoid values $\kp_1\doteq 0$), and consequently also 
the glued matrices $[\bA_{\kp_1}\bH]$ and $[\bA_{\kp_1+\tau}\bH]$ are generically invertible, we can solve $\wt{\bS}$ from \eqref{eq:Nmatsysta} and get 
the equation 
\begin{equation}\label{eq:Seq} 
[\wt{\bG}_{-\kp_1}([\bA_{\kp_1+\tau}\bH]\cdot\bS\cdot[\bA_{\kp_1}\bH]^{-1})]\cdot\bH= [\bA_\tau\bH]\cdot[\bG_{-\kp_1}\bS]\  . 
\end{equation} 
Taking the rank 1 matrix $\bH=\bh^+(\bh^-)^T=(h_i^+h_j^-)$ in the form of a bi-vector, and using the formula for the inverse of 
a Frobenius matrix, \eqref{eq:ACMinv}, this equation can be written explicitly in the following way
\begin{eqnarray}\label{eq:Sbigeq} 
&& \sum_{l,l',l''}\,\Phi_{\kp_1+\tau}(\wt{\xi}_i-\xi_l-\alpha)\,\Phi_{\kp_1}(\tau)\,
\Phi_{\kp_1-\tau}(\xi_{l'}-\wt{\xi}_{l''}+\alpha)\,\Phi_{-\kp_1}(\wt{\xi}_i-\wt{\xi}_{l''}-\wt{\gm}) \nn \\ 
&& \qquad \times \left[\frac{\prod_k\,\sg(\xi_{l'}-\wt{\xi}_k+\alpha)}{\prod_{k\neq l'}\,\sg(\xi_{l'}-\xi_k)} \right]
\left[\frac{\prod_k\,\sg(\wt{\xi}_{l''}-\xi_k-\alpha)}{\prod_{k\neq l''}\,\sg(\wt{\xi}_{l''}-\wt{\xi}_k)} \right]
\left(h_l^-S_{l,l'}(h^-_{l'})^{-1}\right) \nn \\ 
&& = \sum_l\,\Phi_{\tau}(\wt{\xi}_i-\xi_l-\alpha)\,\Phi_{-\kp_1}(\xi_l-\xi_j-\gm)\,\left(h_l^-S_{l,j}(h^-_{j})^{-1}\right)\  , 
\end{eqnarray}
which constitutes a system of $N^2$ homogeneous linear equations for the $N^2$ quantities $h^-_iS_{ij}(h^-_j)^{-1}$ 
involving the entries of the matrix $\bS$. In order for the equation to lead to nontrivial solutions for the latter quantities  
the coefficient determinant must vanish, and this leads to a first-order difference equation in terms of the main dependent variables  
$\xi_j(n)$, which is subject to the additional determinant condition $\det(\bA_\tau)=0$. In the case $N=2$ we are thus led to a first order 
coupled equation for $\xi_1(n)$ and $\xi_2(n)$, but also subject to the 
condition ~$\tau+\xi_1+\xi_2-2\alpha=0$~. Thus, we get effectively a first order equation in terms of a single dependent variable, 
which we expect would again be linearizable. However due to the complexity of the determinant condition following from 
\eqref{eq:Sbigeq} it is hard to do the integration explicitly. Thus, unlike the autonomous case, in order to get nontrivial equations for the 
$\xi_j$ we must proceed to a higher order scheme, which we will do in the next section.

\subsection{Revised scheme} 

We noted from the analysis of the previous subsection that not only the $\kp_1$ parameter becomes irrelevant in the final equations, but also that 
from the rank 1 matrix $\bH=\bh^+(\bh^-)^T$ the dependence on the first factor $\bh^+$ effectively disappears, while from  \eqref{eq:Sbigeq} it 
is apparent that the glued matrix $[\bG_{-\kp_1}\bS]$ can be taken to be of rank 1. Calling the latter matrix $\ol{\bS}=(\ol{S}_{ij})$ we see 
that we can rewrite the Lax matrix \eqref{eq:1storderLaxb} as follows 
\[ (\bT_\kp)_{ij}=S_{ij}\sg(\kp)\,\Phi_\kp(\xi_i-\xi_j-\gm-\kp_1)\,\sg(-\kp_1)\Phi_{-\kp_1}(\xi_i-\xi_j-\gm)
=:\ol{S}_{ij}\sg(\kp)\,\Phi_\kp(\xi_i-\xi_j-\ol{\gm})\ ,  \]
where we have written $\ol{\gm}:=\gm+\kp_1$. Thus, the scheme given by \eqref{eq:1storderLax} is equivalent to one in which for the matrix $\bT_\kp$ we 
take the form
\be\label{eq:1storderLaxc} 
(\bT_{\kp})_{i,j} = \oS_{i,j}\ \sg(\kp)\,\Phi_{\kp}(\xi_i-\xi_j-\ogm)\ ,  
\ee 
instead of the original form \eqref{eq:1storderLaxb}. Working out the compatibility of the system \eqref{eq:compat} we have now the following alternative 
computation to the original one. The left-hand side corresponds to 
\begin{eqnarray*}
&& \sum_{l=1}^N \wt{\oS}_{il}\, H_{lj}\, \Phi_{\kp}(\wt{\xi}_i-\wt{\xi}_l-\wt{\ogm})\Phi_{\kp}(\wt{\xi}_l-\xi_j-\alpha)\ , \\ 
&& = \sum_{l=1}^N \wt{\oS}_{il}\, H_{lj}\, \Phi_{\kp}(\wt{\xi}_i-\xi_j-\alpha-\wt{\ogm})\left[ \zeta(\kp)-\zeta(\kp+\wt{\xi}_i-\xi_j-\alpha-\wt{\ogm}) 
+\zeta(\wt{\xi}_i-\wt{\xi}_l-\wt{\ogm})+\zeta(\wt{\xi}_l-\xi_j-\alpha)\right] 
\end{eqnarray*} 
while the right-hand side yields:
\begin{eqnarray*}
&& \sum_{l=1}^N H_{il}\,\oS_{lj}\,\sg(\tau) \Phi_\kp(\tau)\,\Phi_{\kp+\tau}(\wt{\xi}_i-\xi_l-\alpha)\,\Phi_\kp(\xi_l-\xi_j-\ogm) \\ 
&& = \sum_{l=1}^N H_{il}\,\oS_{lj}\,\sg(\tau) \Phi_\tau(\wt{\xi}_i-\xi_l-\alpha)\,\Phi_{\kp}(\tau+\wt{\xi}_i-\xi_l-\alpha)\,\Phi_\kp(\xi_l-\xi_j-\ogm) \\ 
&& = \sum_{l=1}^N H_{il}\,\oS_{lj}\,\sg(\tau) \Phi_\tau(\wt{\xi}_i-\xi_l-\alpha)\,\Phi_{\kp}(\tau+\wt{\xi}_i-\xi_j-\alpha-\ogm) \times \\ 
&&\qquad \times \left[\zeta(\kp)-\zeta(\kp+\tau+\wt{\xi}_i-\xi_j-\alpha-\ogm)+\zeta(\tau+\wt{\xi}_i-\xi_l-\alpha)+\zeta(\xi_l-\xi_j-\ogm)
\right] 
\end{eqnarray*}
Setting  as before $\wt{\ogm}=\ogm-\tau$, and identifying on both sides the terms that contain $\kp$ and those that don't, we arrive at the system of equations: 

\bse\label{eq:Laxmat3}\bea
&& \sum_{l=1}^N \wt{\oS}_{il} H_{lj}=\sum_{l=1}^N H_{il}\oS_{lj}\,\sg(\tau)\,\Phi_{\tau}(\wt{\xi}_i-\xi_l-\alpha)\ , \label{eq:Laxmat3a}\\
&& \sum_{l=1}^N \wt{\oS}_{il} H_{lj}\,\sg(-\tau)\,\Phi_{-\tau}(\wt{\xi}_i-\wt{\xi_l}-\wt{\ogm})\,\Phi_{-\tau}(\wt{\xi}_l-\xi_j-\alpha)=
\sum_{l=1}^N H_{il}\oS_{lj}\,\Phi_{-\tau}(\xi_l-\xi_j-\ogm), \nn \\ 
&& \qquad \qquad i,j=1,\dots,N\ , \label{eq:Laxmat3b}  
\eea\ese
which is essentially equivalent to \eqref{eq:1stordsyst}, except that the system is written in terms of 
variables $\oS_{ij}$ rather than $S_{ij}$. Since the latter quantities were to be determined from the Lax 
compatibility in the first place, where now the parameter $\kp_1$ is absorbed in the yet to be 
determined variables, it is clear that the final equations for $\xi$ will not involve the latter parameter. 
We note also that the way in which the second relation \eqref{eq:Laxmat3b} was obtained, was achieved by 
extracting the the terms containing the $\zeta$ functions in the compatibility, and then recombining them 
with the appropriate terms from the first relation \eqref{eq:Laxmat3a} in such a way that we essentially get 
the full Lax compatibility relation back from which we started but for any fixed value of the spectral 
parameter, $\kp=\kp_0$ say. Since we only need one such relation together with \eqref{eq:Laxmat3a} to have 
the full set of compatibility relations, it suffices to set $\kp_0=-\tau$ in order to obtain 
\eqref{eq:Laxmat3b}. In matrix form, using the gluing convention introduced earlier, the resulting 
system can be written conveniently as follows: 
\bse\label{eq:revmatsyst}\begin{eqnarray}
&& \wt{\obS}\cdot \bH = [\bA_\tau\bH]\cdot \obS,  \label{eq:revmatsysta} \\
&& [\wt{\obG}_{-\tau}\wt{\ol{\bS}}]\cdot[\bA_{-\tau}\bH] = \bH \cdot [\obG_{-\tau}\ol{\bS}]\ , 
\label{eq:revmatsystb}
\end{eqnarray} \ese 
where $\obG_\kp$ denotes the matrix $\bG_\kp$ in which $\gm$ is replaced by $\ogm$. 
  
We will now analyse the system \eqref{eq:revmatsyst}, which for generic $\kp_1$ is equivalent\footnote{In fact, from the 
Lax equation \eqref{eq:compat} for \eqref{eq:1storderLax}, we have that ~$[\wt{\bG}_{\kp-\kp_1}\wt{\bS}]\cdot[\bA_\kp\bH]=
[\bA_{\kp+\tau}\bH]\cdot[\bG_{\kp-\kp_1}\bS]$~ for arbitrary fixed $\kp$, and this will lead to either system \eqref{eq:Nmatsyst} 
or \eqref{eq:revmatsyst} with suitable choices for $\kp$.}  to \eqref{eq:Nmatsyst}. 
First, it is easily derived from the explicit form \eqref{eq:Laxmat3}, and by using the relation \eqref{eq:16} of Appendix B, that for $N=1$ 
we obtain once again a linearizable system system for $\xi(n_:=\xi_1(n)$, namely  
\[ \wp(\wt{\xi}-\xi-\alpha)=\wp(\ogm) \quad \Rightarrow \quad  \wt{\xi}-\xi-\alpha \doteq \pm \ogm\  , \] 
leading to the solution 
\[  \xi(n)=\xi(0)+ (\alpha\pm\ogm(0)+\Omega)n\pm \tfrac{1}{2}n(n-1)\tau\  .  \] 

When $N=2$, taking as before the matrix $\bH$ to be of rank 1, there are two possibilities: from \eqref{eq:revmatsysta} either, \emph{i)} $\obS$ is of rank 1,
or, \emph{ii)} $[\bA_\tau\bH]$ is of rank 1, implying that $\det(\bA_\tau)=0$. In case \emph{i)} we would conclude that $\det(\bA_{-\tau})=0$ (since 
otherwise $\det(\obG_{-\tau})=0$ and this would lead to special conditions on $\gm$), while in case \emph{ii)} we would conclude that $[\obG_{-\tau}\obS]$ 
is of rank 1. Both options lead to similar results, so for convenience let is pursue the case \emph{i)}. In hat case we have the condition: 
\[ \det(\bA_{-\tau})=0 \quad  \Rightarrow \quad -\tau+\wt{\Xi}-\Xi-2\alpha\doteq 0\  , \] 
for ~$\Xi=\xi_1+\xi_2$~, as follows from the Frobenius determinant formula \eqref{eq:cauchy}.

To resolve this case \emph{i)}, let us write once again $\bH=\bh^+(\bh^-)^T$ and $\obS=\bss^+(\bss^-)^T$, with entries $\oS_{ij}=s^+_i s^-_j$, then the 
the first relation \eqref{eq:revmatsysta} allows us to identify $s^-_j=\rho h^-_j$ (for some scalar function $\rho$), and consequently:
\[
(\wt{\bss}^-\cdot\bh^+)\frac{\wt{s}_i^+}{h_i^+}=\sum_{l=1}^2\,(\bA_\tau)_{il}s_l^-s_l^+\   . 
\] 
The second relation \eqref{eq:revmatsystb} leads to the condition
\[ 
\frac{\wt{s}_i^+}{h_i^+}\,\sum_{l=1}^2(\wt{\obG}_{-\tau})_{il}\wt{s}^-_lh_l^+(\bA_{-\tau})_{lj}=\sum_{l=1}^2s_l^-s_l^+(\obG_{-\tau})_{lj}\  , 
\] 
Expressing all the entries of the first and second relation in terms of $s^-_l s^+_l=:S_l$,\, $\wt{s}^-_l h^+_l=:H_l$ we get a system of equations 
comprising on the one hand  
\bse\label{eq:firstreducedsystem}\bea
&&\Big(1+ \frac{H_2}{H_1}\Big)\wt{S}_1 = A^+_{11} S_1 +A^+_{12} S_2\ ,\\
&&\Big(\frac{H_1}{H_2}+1 \Big)\wt{S}_2 = A^+_{21} S_1 +A^+_{22} S_2\ ,
\eea\ese
where we abbreviated $A^\pm_{ij}:=(\bA_{\pm\tau})_{ij}$, and on the other hand, with $G^{\pm}_{ij}:=(\obG_{\pm\tau})_{ij}$,  
\begin{eqnarray}\label{eq:reducdsystm}
&& \left(\wt{G}^-_{11}A^-_{11} + \wt{G}^-_{12}A^-_{21} \frac{H_2}{H_1}\right) \wt{S}_1 = \left( \wt{G}^-_{21}A^-_{11} \frac{H_1}{H_2} + \wt{G}^-_{22}A^-_{21} \right) \wt{S}_2 = 
G^-_{11} S_1 + G^-_{21} S_2\ , \nn\\
&& \left(\wt{G}^-_{11}A^-_{12} + \wt{G}^-_{12}A^-_{22} \frac{H_2}{H_1}\right) \wt{S}_1 = \left( \wt{G}^-_{21}A^-_{12} \frac{H_1}{H_2} + \wt{G}^-_{22}A^-_{22} \right) \wt{S}_2 = 
G^-_{12} S_1 + G^-_{22} S_2\ . \nn\\
\end{eqnarray}
Rewriting these relations in terms of $X=H_2/H_1$ and $Y=S_2/S_1$, we are led to: 
\bea\label{eq:YZX}
&&\frac{\wt{Y}}{X} = \frac{A^+_{21}+A^+_{22}Y}{A^+_{11}+A^+_{12}Y}=\frac{\wt{G}^-_{11}A^-_{11}+\wt{G}^-_{12}A^-_{21}X}{\wt{G}^-_{21}A^-_{11}+\wt{G}^-_{22}A^-_{21}X}\ ,\nn \\
&& (1+X)(G^-_{11}+G^-_{21}Y)=(A^+_{11}+A^+_{12}Y)(\wt{G}^-_{11}A^-_{11}+\wt{G}^-_{12}A^-_{21}X)\ , \nn \\
&&A^-_{12}/A^-_{11}=A^-_{22}/A^-_{21}=\frac{G^-_{12}+G^-_{22}Y}{G^-_{11}+G^-_{21}Y}\ .
\eea
These constitute, in fact, a system four independent relations for $X$, $Y$ and $\wt{Y}$ with coefficients in terms of $\xi_1$ and $\xi_2$, and can be solved by direct computation. 
Eliminating $X$, $Y$ and $\wt{Y}$ we get a rather complicated first order difference equation in terms of $\xi_1$ and $\xi_2$, which we refrain from writing down here, and which is subject to the relation $-\tau+\xi_1+\xi_2-2\alpha+\doteq 0$.  Because of the latter condition we expect the resulting equation for one of the variables, say 
$\xi_1$, and containing three free parameters, $\alpha$, $\Xi(0)$ and $\gm_0$ (apart from the step size $\tau$), to be linearisable, but we have not yet done so. 

Although the first order scheme described in this section and the first order elliptic difference equations resulting from them may be interesting in their own right, 
the scheme seems not yet rich enough to obtain higher order O$\Delta$Es, e.g. elliptic difference equations of Painlev\'e type. This is as expected, as the 
parallel with the  monodromy problem for P$_{\text{VI}}$ and its connection with lattice KdV systems, cf. \cite{NRGO}, indicates that we need at least two lattice 
directions to get interesting transcendental equations. Thus, we will next investigate the structure of the higher order elliptic scheme.

\section{Higher order revised scheme}

We noted in the previous section that the monodromy part of the Lax pair, i.e., \eqref{eq:1storderLaxb}, can be simplified by redefining 
the coefficient matrix $S_{ij}$ and the moving parameter $\gm$, absorbing the parameter $\kp_1$. This applies also to the 
general discrete monodromy problem \eqref{eq:Laxxmatricesssb}, which accordingly can be simplified to the following form: 
\be\label{eq:revisedLaseq} 
(\bT_{\kp})_{i,j} = \sum_{l_1,\dots,l_{m-1}}\,\oS_{i,j}^{(l_1,\dots,l_{m-1})}\,\prod_{\nu=1}^m 
 \sg(\kp)\,\Phi_{\kp}(\xi^{(\nu-1)}_{l_{\nu-1}}-\xi^{(\nu)}_{l_\nu}-\ogm_\nu)\   ,  
\ee 
by redefining
\[ \oS_{i,j}^{(l_1,\dots,l_{m-1})}:= S_{i,j}^{(l_1,\dots,l_{m-1})}\prod_{\nu=1}^m\,\sg(-\kp_\nu)\,
\Phi_{-\kp_\nu}(\xi^{(\nu-1)}_{l_{\nu-1}}-\xi^{(\nu)}_{l_\nu}-\gm_\nu)\  , 
\quad \ogm_\nu:=\gm_\nu+\kp_\nu\  .   \]
The latter redefinition is possible since \emph{ab initio} we don't specify the coefficient matrices, requiring them only to be independent of the spectral parameter 
$\kp$. All further properties of the coefficients should follow from the consistency conditions and additional natural choices (e.g., imposing a rank 1 condition 
on the coefficient matrix in \eqref{eq:Laxxmatricesssa} in accordance with the choices of section 2).  We now reexamine the consistency condition \eqref{eq:compat} 
of \eqref{eq:Laxxmatricesssa} and \eqref{eq:revisedLaseq} for the higher order case. 

\subsection{Second order scheme} 

As is clear from the first order case, treated in subsection 3.3, this requires functional identities for products of the form 
$\prod_{\nu}\,\Phi_\kp(x_\nu)$, i.e. of functions $\Phi_\kp$ for the same value of the label.  Such identities generalizing \eqref{eq:12}, which is equivalent 
to one of the standard addition formulae for Weierstrass functions, were discussed in Appendix C, cf. \eqref{eq:tripleprod} and \eqref{eq:quadrupleprod}, 
and the generalized form is described in the Lemma of the Appendix. The key feature of these higher order elliptic addition formulae is that 
they allow us to separate the spectral parameter dependence of the Lax compatibility conditions, and thus to derive a set of basic relations from 
which the coefficient matrices can be eliminated. A general closed-form formula is hard to give (Lemma 1 gives us a general prescription) for arbitrary 
orders, in contrast to the original scheme presented in section 3. However, the latter contains redundant parameters which are hard to get rid of 
in higher orders, when trying to capture what goes on in concrete formulae. Thus, we will restrict ourselves in this section to the case of a second 
order scheme  (i.e. $m=2$), which conveys adequately the ideas. The corresponding monodromy problem in revised form, in that case is given by:    
\begin{eqnarray}\label{eq:higherorderredc}
 \bchi_{\kp+\tau}&=&\bT'_\kp\,\bchi_\kp\  , \\ 
(\bT'_\kp)_{i,j} &:=&  \sg^2(\kp)\,\sum_{l'=1}^N S_{i,j}^{(l')}\Phi_{\kp}(\xi_i-\eta_{l'})\Phi_\kp(\eta_{l'}-\xi_j-\gamma)\  ,\quad  (i,j=1,\dots,N)\  ,  \nn 
\end{eqnarray}  
where, for notational convenience, we have omitted the $\ol{\phantom{}}$ notation and renamed the $\xi_{l_1}^{(1)}+\ogm_1=:\eta_{l'}$, denoting $l_1=l'$. 
The coefficient variables $S_{i,j}^{(l')}$ remain to be determined when we consider this difference equation on the torus in conjunction with the lattice Lax 
system given by 
\bse\label{eq:higherorderredcLax}\begin{eqnarray}
&& \wt{\bchi}_\kp =\bL_\kp\,\bchi_\kp\  ,  \quad  (\bL_\kp)_{i,j} = H_{i,j}\ \sg(\kp) \,\Phi_{\kp}(\wt{\xi}_i-\xi_j-\alpha)\ , \label{eq:higherorderredcLaxa} \\ 
&& \wh{\bchi}_\kp =\bM_\kp\,\bchi_\kp\  ,   \quad (\bM_\kp)_{i,j} = K_{i,j}\ \sg(\kp) \,\Phi_{\kp}(\wh{\xi}_i-\xi_j-\beta)\  , \label{eq:higherorderredcLaxb}
\end{eqnarray}\ese
which constitutes a system of the type considered in section 2, but without prejudice for now about the form of the coefficient matrices 
$\bH=(H_{ij})$ and $\bK=(K_{ij})$, but which we will assume in due course to be of rank 1 in accordance with the derivation in section 2. The only further 
assumption on the latter is that they are independent of the spectral parameter $\kp$.   

To give a motivation of the monodromy problem associated with \eqref{eq:higherorderredc}, we consider the variable $\eta$ as an intermediate dependent variable 
on a 2-step configuration in a multidimensional lattice, where the elementary shifts correspond to Lax operations of the type \eqref{eq:higherorderredcLax} 
but in perhaps additional lattice directions. Thus, the monodromy problem given in the form of the elliptic difference equation \eqref{eq:higherorderredc} 
would amount to a de-autonomization  of a 2-step periodic reduction on that lattice. The 2-step periodic reduction, illustrated in the diagram below, 
leads to a spectral problem of the form: ~$\wh{\ol{\bchi}}=\ld\bchi$, constituting the spectral part of a Lax pair describing a higher-order stationary discrete 
flow. A subsequent de-autonomization, in the spirit of the paper \cite{PNGR}, on the level of the Lax 
representation by making the replacement ~$\ld\bchi\ \ \leadsto\  \bchi_{\kp+\tau}$~, leads then to a monodromy problem of the form \eqref{eq:higherorderredc}, 
in which the intermediate value $\eta$ of the dependent variable $\xi$, shifted in a relevant direction, can be left unspecified. This allows us 
to determine the value $\eta$ from the consistency of the Lax pair.  

\begin{figure}[ht]
\centering
\setlength{\unitlength}{0.0003in}
\begingroup\makeatletter\ifx\SetFigFont\undefined%
\gdef\SetFigFont#1#2#3#4#5{%
  \reset@font\fontsize{#1}{#2pt}%
  \fontfamily{#3}\fontseries{#4}\fontshape{#5}%
  \selectfont}%
\fi\endgroup%
{\renewcommand{\dashlinestretch}{30}
\begin{picture}(4182,3705)(0,-10)
\put(652,3218){\circle*{144}}
\put(3420,3195){\circle*{144}}
\put(3446,508){\circle*{144}}
\thicklines
\drawline(675,3195)(3240,3195)
\drawline(675,3195)(3240,3195)
\drawline(3000.000,3135.000)(3240.000,3195.000)(3000.000,3255.000)(3000.000,3135.000)
\drawline(3420,3150)(3420,675)
\drawline(3420,3150)(3420,675)
\drawline(3360.000,915.000)(3420.000,675.000)(3480.000,915.000)(3360.000,915.000)
\put(495,3510){\makebox(0,0)[lb]{$\bchi,\bxi$}}
\put(3420,3465){\makebox(0,0)[lb]{$\overline{\bchi},\boldsymbol \eta$}}
\put(3690,270){\makebox(0,0)[lb]{$\ld\bchi,\bxi$}}
\end{picture}
\caption{2-step periodic reduction.}\label{2-step_reduc}
\label{2-step periodic}
}
\end{figure}
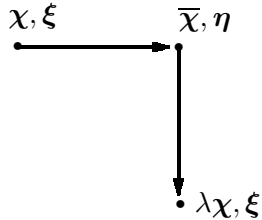

The elliptic isomonodromic deformation system comprising \eqref{eq:higherorderredc} and \eqref{eq:higherorderredcLax}   
leads to the following set of compatibility conditions: 
\bse\label{eq:highercasecompatibility}\begin{eqnarray} 
\wt{\bT}'_\kp\,\bL_\kp &=& \bL_{\kp+\tau}\,\bT'_\kp\ ,\label{eq:highercasecompatibilitya}\\
\wh{\bT}'_\kp\,\bM_\kp &=& \bM_{\kp+\tau}\,\bT'_\kp\ ,\label{eq:highercasecompatibilityb}\\ 
\wh{\bL}_\kp\,\bM_\kp &=& \wt{\bM}_\kp\,\bL_\kp\ .  \label{eq:highercasecompatibilityc}
\end{eqnarray} \ese
We note that the combination of multiple lattice shifts appearing in the monodromy problems of Painlev\'e type 
equations was first made apparent in \cite{NRGO,NW} where it was made manifest in the monodromy problem of P$_{\rm VI}$, 
and in the structure of lattice Garnier systems. Recently, in \cite{OrmerodRains}, this was also 
exploited in the construction of $q$-difference Garnier systems. In what follows, we will focus on one the 
lattice shifts, namely the one generated by \eqref{eq:higherorderredcLaxa}, and the compatibility 
condition \eqref{eq:highercasecompatibilitya}, while \eqref{eq:highercasecompatibilityc} was analysed in section 2. 
  
To analyse the system \eqref{eq:highercasecompatibilitya} most effectively we need to use the (seemingly novel) 
elliptic identities presented in Appendix C, the structure of which is summarised in the Lemma. In particular, 
where we used \eqref{eq:12} in the first order case, we now need identities such as \eqref{eq:tripleprod} 
and \eqref{eq:quadrupleprod} to analyse the structure of the compatibility relations.  
The consistency condition \eqref{eq:highercasecompatibilitya} can be worked out as follows. The left-hand side can be 
written as 
\begin{eqnarray*} 
&& \sum_{l,l'=1}^N \wt{S}_{il}^{(l')}\,H_{lj}\Phi_\kp(\wt{\xi}_i-\wt{\eta}_{l'})\,\Phi_\kp(\wt{\eta}_{l'}-\wt{\xi}_l-\wt{\gamma})\,
\Phi_\kp(\wt{\xi}_l-\xi_j-\alpha)  \\ 
&& = \sum_{l,l'=1}^N \wt{S}_{il}^{(l')}\,H_{lj}\,\tfrac{1}{2}\Phi_\kp(\wt{\xi}_i-\xi_j-\alpha-\wt{\gamma})\times \\
&&\quad\times\Big[ \Big( \zeta(\kp)-\zeta(\kp+\wt{\xi}_i-\xi_j-\alpha-\wt{\gamma})+
\zeta(\wt{\xi}_i-\wt{\eta}_{l'})+\zeta(\wt{\eta}_{l'}-\wt{\xi}_l-\wt{\gamma})+\zeta(\wt{\xi}_l-\xi_j-\alpha) \Big)^2 \\  
&& \qquad +\wp(\kp)-\Big(\wp(\kp+\wt{\xi}_i-\xi_j-\alpha-\wt{\gamma})+\wp(\wt{\xi}_i-\wt{\eta}_{l'})+
\wp(\wt{\eta}_{l'}-\wt{\xi}_l-\wt{\gamma})+
\wp(\wt{\xi}_{l}-\xi_j-\alpha) \Big)\Big] \ ,  
\end{eqnarray*} 
where we have made use of the identity \eqref{eq:tripleprod}. The right-hand side, using again the identity $\Phi_\kp(\tau)\,\Phi_{\kp+\tau}(x)=\Phi_\tau(x)\,\Phi_\kp(\tau+x)$ 
leads to 
\begin{eqnarray*}
&&\sum_{l,l'=1}^N H_{il}\,S_{lj}^{(l')}\,\sg(\tau)\Phi_{\kp}(\tau)\,\Phi_{\kp+\tau}(\wt{\xi}_i-\xi_{l}-\alpha)\,
\Phi_\kp(\xi_l-\eta_{l'})\,\Phi_\kp(\eta_{l'}-\xi_j-\gamma) \\ 
&& = \sum_{l,l'=1}^N H_{il}\,S_{lj}^{(l')}\,\sg(\tau)\,\Phi_{\tau}(\wt{\xi}_i-\xi_{l}-\alpha)\,\Phi_{\kp}(\tau+\wt{\xi}_i-\xi_{l}-\alpha)\,
\Phi_\kp(\xi_l-\eta_{l'})\,\Phi_\kp(\eta_{l'}-\xi_j-\gamma)  \\
&&=\sum_{l,l'=1}^N H_{il}\, S_{lj}^{(l')}\,\tfrac{1}{2} 
\sg(\tau)\Phi_\tau(\wt{\xi}_i-\xi_l-\alpha)\Phi_\kp(\wt{\xi}_i-\xi_j-\alpha-\gamma+\tau)\times \\ 
&&\qquad \times \Big[ \Big( \zeta(\kp)-\zeta(\kp+\wt{\xi}_i-\xi_j-\alpha-\gamma+\tau)+\zeta(\tau+\wt{\xi}_i-\xi_l-\alpha)+\zeta(\xi_l-\eta_{l'})+
\zeta(\eta_{l'}-\xi_j-\gamma) \Big)^2 \\  
&& \qquad +\wp(\kp)-\Big(\wp(\kp+\wt{\xi}_i-\xi_j-\alpha-\gamma+\tau)+\wp(\tau+\wt{\xi}_i-\xi_l-\alpha)+\wp(\xi_l-\eta_{l'})+
\wp(\eta_{l'}-\xi_j-\gamma) \Big)\Big] \ . 
\end{eqnarray*} 
Setting (once again) ~$\wt{\gamma}=\gamma-\tau$~ the common factor 
$\Phi_\kp(\wt{\xi}_i-\xi_j-\alpha-\wt{\gamma})=\Phi_\kp(\wt{\xi}_i-\xi_j-\alpha-\gamma+\tau)$ on both sides cancel out, and after factoring them 
out the remaining terms can be separated in accordance with their different dependence on $\kp$. 
The latter only appears in combination with the external indices $i,j$ and do not mix with the summation indices. Thus, we only have three types 
of terms w.r.t. to the dependence on $\kp$: constant terms, terms linear in  $(\zeta(\kp)-\zeta(\kp+\wt{\xi}_i-\xi_j-\alpha-\wt{\gamma})$ and 
terms of the form $(\zeta(\kp)-\zeta(\kp+\wt{\xi}_i-\xi_j-\alpha-\gamma+\tau)^2+\wp(\kp)-\Big(\wp(\kp+\wt{\xi}_i-\xi_j-\alpha-\gamma+\tau)$. 
These then yield the following relations (in reverse order): 
\bse\label{eq:firsthc}\begin{eqnarray}\label{eq:firsthca}
&& \sum_{l,l'=1}^N \wt{S}_{il}^{(l')}\,H_{lj}=\sum_{l,l'=1}^N \sg(\tau)\Phi_\tau(\wt{\xi}_i-\xi_l-\alpha)\,H_{il}\,S_{lj}^{(l')}\ , \\ 
&& \sum_{l,l'=1}^N \wt{S}_{il}^{(l')}\,H_{lj}\,
\Big[\zeta(\wt{\xi}_i-\wt{\eta}_{l'})+\zeta(\wt{\eta}_{l'}-\wt{\xi}_l-\gamma+\tau)+\zeta(\wt{\xi}_l-\xi_j-\alpha) \Big]\nn \\
&&=\sum_{l,l'=1}^N \sg(\tau)\Phi_\tau(\wt{\xi}_i-\xi_l-\alpha)\,H_{il}\,S_{lj}^{(l')}
\Big[ \zeta(\tau+\wt{\xi}_i-\xi_l-\alpha)+\zeta(\xi_l-\eta_{l'})+\zeta(\eta_{l'}-\xi_j-\gamma) \Big]\ ,  \nn\\ 
&& \label{eq:firsthcb} 
\end{eqnarray} 
and 
\begin{eqnarray} 
&& \sum_{l,l'=1}^N \wt{S}_{il}^{(l')}\,H_{lj}\,\Big[\Big(\zeta(\wt{\xi}_i-\wt{\eta}_{l'})+\zeta(\wt{\eta}_{l'}-\wt{\xi}_l-\gamma+\tau)+\zeta(\wt{\xi}_l-\xi_j-\alpha) \Big)^2\nn \\
&&\hspace{4cm}-\wp(\wt{\xi}_i-\wt{\eta}_{l'})-\wp(\wt{\eta}_{l'}-\wt{\xi}_l-\gamma+\tau)-\wp(\wt{\xi}_{l}-\xi_j-\alpha) \Big] \nn  \\ 
&&=\sum_{l,l'=1}^N H_{il}\,S_{lj}^{(l')}\,\Phi_{\tau}(\wt{\xi}_i-\xi_l-\alpha)\,\sg(\tau)\Big[\Big( 
\zeta(\tau+\wt{\xi}_i-\xi_l-\alpha)+\zeta(\xi_l-\eta_{l'})+\zeta(\eta_{l'}-\xi_j-\gamma)\Big)^2\nn\\
&&\hspace{4cm}-\wp(\tau+\wt{\xi}_i-\xi_l-\alpha)-\wp(\xi_l-\eta_{l'})-\wp(\eta_{l'}-\xi_j-\gamma)\Big] \label{eq:firsthcc}
\end{eqnarray}\ese 
By combining \eqref{eq:firsthcb} with \eqref{eq:firsthca}, and using \eqref{eq:zs}, the former can also be re-cast in the form 
\begin{eqnarray}  
&& \sum_{l,l'=1}^N \wt{S}_{il}^{(l')}\,H_{lj}\,
\frac{\sg(\wt{\xi}_i-\wt{\xi}_l-\wt{\gamma})\sg(\wt{\eta}_{l'}-\xi_j-\wt{\gamma}-\alpha)\sg(\wt{\xi}_i+\wt{\xi}_l-\xi_j-\wt{\eta}_{l'}-\alpha)}{\sg(\wt{\xi}_i-\wt{\eta}_{l'})\sg(\wt{\eta}_{l'}-\wt{\xi}_l-\wt{\gamma})\sg(\wt{\xi}_l-\xi_j-\alpha)}   \nn \\ 
&& = \sum_{l,l'=1}^N H_{il}\,S_{lj}^{(l')}\,\frac{\sg(\wt{\xi}_i-\eta_{l'}+\tau-\alpha)\sg(\xi_l-\xi_j-\gamma)\sg(\wt{\xi}_i-\xi_l-\xi_j+\eta_{l'}-\alpha-\wt{\gamma})}
{\sg(\wt{\xi}_i-\xi_l-\alpha+\tau)\sg(\xi_l-\eta_{l'})\sg(\eta_{l'}-\xi_j-\gamma)}\  , \label{eq:secondhc} 
\end{eqnarray} 
while combining \eqref{eq:firsthcc} with both \eqref{eq:firsthcb} and \eqref{eq:firsthca}, and using again \eqref{eq:tripleprod}, we can obtain a 
simpler form of the third relation, namely 
\begin{eqnarray} 
&& \sum_{l,l'=1}^N \wt{S}_{il}^{(l')}\,H_{lj}\,\sg(-\tau)\,\Phi_{-\tau}(\wt{\xi}_i-\wt{\eta}_{l'})\,\Phi_{-\tau}(\wt{\eta}_{l'}-\wt{\xi}_l-\wt{\gamma})
\,\Phi_{-\tau}(\wt{\xi}_l-\xi_j-\alpha) \nn \\ 
&& =\sum_{l,l'=1}^N H_{il}\,S_{lj}^{(l')}\,\Phi_{-\tau}(\xi_l-\eta_{l'})\,\Phi_{-\tau}(\eta_{l'}-\xi_j-\gamma)\  . \label{eq:thirdhc} 
\end{eqnarray} 
As before, we want to eliminate the quantities $S^{(l')}_{i,j}$ and $H_{i,j}$ in these relations to obtain a (possibly coupled) system of equations for the 
variables $\xi_j$ and $\eta_j$ alone. To do the analysis it may prove helpful to use the notation introduced  in section 3, using glued matrices. In that notation 
eqs. \eqref{eq:firsthca} and \eqref{eq:thirdhc} can be written as 
\bse\label{rel:firstlast}\begin{eqnarray}
&& [\wt{\bS}]\cdot\bH = [\bA_\tau\bH]\cdot[\bS]\ , \label{rel:firstlasta}\\ 
&& [\wt{\boldsymbol{\bE}}_{-\tau}[\wt{\bS}]\wt{\bF}_{-\tau}]\cdot[\bA_{-\tau}\bH]=
\bH\cdot[\bE_{-\tau}[\bS]\bF_{-\tau}]\   \label{rel:firstlastb}, 
\end{eqnarray} \ese
with $\bA_{\pm\tau}$ as given in section 3, and where $[\bS]$ denotes the matrix with entries $([\bS])_{i,j}=\sum_{l'}S_{ij}^{(l')}$. Furthermore, we are compelled to 
introduce a somewhat ad-hoc notation for  
\[   \sum_{l'} S_{ij}^{(l')}\,\sg^2(-\tau)\Phi_{-\tau}(\xi_i-\eta_{l'})\,\Phi_{-\tau}(\eta_{l'}-\xi_j-\gamma)=:[\bE_{-\tau}[\bS]\bG_{-\tau}]_{ij}\   \] 
a kind of `doubly glued' matrix involving the upper index in the quantity $S_{i,j}^{(l')}$, and where $\bE_{\kp}$ and $\bF_\kp$ (for arbitrary $\kp$) denote the matrices with entries 
\[ (\bE_\kp)_{i,j}:=\sg(\kp)\,\Phi_\kp(\xi_i-\eta_j)\  , \quad (\bF_\kp)_{ij}:=\sg(\kp)\,\Phi_\kp(\eta_i-\xi_j-\gm) . \] 

The `middle' relation \eqref{eq:secondhc}, or (modulo the first relation) equivalently \eqref{eq:firsthcb}, is the more complicated one to write in matrix form. 
To achieve that we rewrite the original form \eqref{eq:firsthcb} as follows: 
\begin{eqnarray*} 
&&\sum_{l,l'=1}^N \wt{S}_{il}^{(l')}\,H_{lj}\,
\Big[\zeta(\wt{\xi}_i-\wt{\eta}_{l'})+\zeta(\wt{\eta}_{l'}-\wt{\xi}_l-\gamma+\tau)+\zeta(-\tau)-\zeta(\wt{\xi}_{i}-\wt{\xi}_l-\gamma) \\
&&\qquad +\zeta(\wt{\xi}_{i}-\wt{\xi}_l-\gamma)+\zeta(\tau)+\zeta(\wt{\xi}_l-\xi_j-\alpha) -\zeta(\wt{\xi}_i-\xi_j-\alpha-\wt{\gamma})\Big] \\ 
&&=\sum_{l,l'=1}^N H_{il}\,S_{lj}^{(l')} \sg(\tau)\Phi_\tau(\wt{\xi}_i-\xi_l-\alpha)\,
\Big[ \zeta(\tau+\wt{\xi}_i-\xi_l-\alpha)+\zeta(\xi_l-\xi_j-\wt{\gamma})+\zeta(-\tau)\nn\\
&&\qquad-\zeta(\wt{\xi}_i-\xi_j-\alpha-\wt{\gamma})+\zeta(\xi_l-\eta_{l'})+\zeta(\eta_{l'}-\xi_j-\gamma)+\zeta(\tau)-\zeta(\xi_l-\xi_j-\wt{\gamma}) \Big]\  ,  
\end{eqnarray*}
and apply the identity \eqref{eq:12} on each quadruple of $\zeta$-terms in the summands, thus, obtaining:  
\begin{eqnarray}\label{id:linearinkappamatrixrev}
&& \sum_{l,l'=1}^N \frac{\wt{S}_{il}^{(l')}}{\Phi_{-\tau}(\wt{\xi}_i-\wt{\xi}_l-\wt{\gamma})}\,H_{lj}\,
\Big[\Phi_{-\tau}(\wt{\xi}_i-\wt{\eta}_{l'})\,\Phi_{-\tau}(\wt{\eta}_{l'}-\wt{\xi}_l-\wt{\gamma})+
\Phi_{-\tau}(\wt{\xi}_i-\xi_j-\alpha-\wt{\gamma})\,\Phi_{\tau}(\wt{\xi}_l-\xi_j-\alpha)\Big] \nn \\ 
&& =\sum_{l,l'=1}^N H_{il}\,\frac{S_{lj}^{(l')}}{\Phi_\tau(\xi_l-\xi_j-\gamma)} \sg(\tau)\,\Phi_\tau(\wt{\xi}-\xi_l-\alpha) \nn \\ 
&& \qquad\qquad\times  \Big[ \Phi_\tau(\wt{\xi}-\xi_j-\alpha-\wt{\gm})\,\Phi_{-\tau}(\tau+\wt{\xi}_i-\xi_l-\alpha)
+\Phi_\tau(\xi_l-\eta_{l'})\,\Phi_\tau(\eta_{l'}-\xi_j-\gm)\Big]  
\end{eqnarray}
This relation can be written more concisely using the notation of glued matrices, extending the latter further by introducing the notation
\[ ([\bA/\bB])_{ij}:=A_{ij}/B_{ij}\  , \] 
in the following form 
\begin{eqnarray}\label{eq:matrixhigher}
&& \big[\,[\wt{\bE}_{-\tau}[\wt{\bS}]\wt{\bF}_{-\tau}]/\wt{\bG}_{-\tau}\big]\cdot\bH - \big[\left( [\wt{\bS}/\wt{\bG}_{-\tau}]\cdot[\bA_\tau\bH]\right)\bC_{-\tau}\big] \nn \\ 
&& =    \big[\left( \bH\cdot[\bS/\bG_\tau]\right)\bC_\tau\big] - [\bA_\tau\bH]\cdot\big[ [\bE_\tau[\bS]\bF]/\bG_\tau\big]\  , 
\end{eqnarray} 
where $\bC_\kp$ (for arbitrary $\kp$) is given by  
\[ 
(\bC_\kp)_{ij}=\sigma(\kp)\,\Phi_\kp(\wt{\xi}_i-\xi_j-\alpha-\wt{\gamma})\ .
\]
In spite of the unconventional notation, we believe this way of writing the relations to which the coefficients 
are subject are somewhat more insightful than the expressions in terms of components. 

\subsection{Case $N=1$}

For $N=1$ the original system of equations \eqref{eq:firsthc} takes a simple form in terms of the single variables $\xi:=\xi_1$, $\eta:=\eta_1$, and 
leads to the coupled system of equations: 
\bse \label{eq:HigherN=1}\begin{eqnarray}
&& \zeta(\wt{\xi}-\wt{\eta})+ \zeta(\wt{\eta}-\wt{\xi}-\wt{\gamma})+\zeta(\wt{\xi}-\xi-\alpha)
=\zeta(\wt{\xi}-\xi+\tau-\alpha)+\zeta(\xi-\eta)+\zeta(\eta-\xi-\gamma)\  , \nn  \\
&& \\  
&& \wp(\wt{\xi}-\wt{\eta})+ \wp(\wt{\eta}-\wt{\xi}-\wt{\gamma})+\wp(\wt{\xi}-\xi-\alpha)
=\wp(\wt{\xi}-\xi-\alpha+\tau)+\wp(\xi-\eta)+\wp(\eta-\xi-\gamma)\  , \nn \\ 
&&  
\end{eqnarray}\ese 
for $\xi$ and $\eta$, together with the relation 
\[ \frac{\wt{S}}{S}=\sg(\tau)\Phi_{\tau}(\wt{\xi}-\xi-\alpha)\  , \]
for $S:=S_{1,1}$. There are various solutions of the system \eqref{eq:HigherN=1} via the viable identifications of the terms, namely \\ 
\bse\label{eq:N=1sols}
\emph{Case i)} 
\begin{eqnarray}\label{eq:N=1solsa}
&&\left.\begin{array}{rcl}
\wt{\xi}-\xi-\alpha & \doteq &\eta-\xi-\gm \\ 
\wt{\xi}-\wt{\eta} & \doteq & \xi-\eta \end{array}\right\} \quad \Rightarrow\quad \nn \\ 
&& \left\{ \begin{array}{rcl} 
\xi(n) &=& \xi(0)+\left(\eta(0)-\xi(0)-\gm(0)+\alpha+\Omega\right)n+\tfrac{1}{2}n(n-1)(\tau+\Omega') \\ 
\eta(n) &=& \eta(0)+\left(\eta(0)-\xi(0)-\gm(0)+\alpha+\Omega\right)n+\tfrac{1}{2}n(n-1)(\tau+\Omega') \end{array}\right. 
\end{eqnarray}
\emph{Case ii)} 
\begin{eqnarray}\label{eq:N=1solsb}
&&\left.\begin{array}{rcl}
\wt{\xi}-\xi-\alpha & \doteq &\xi-\eta \\ 
\wt{\eta}-\wt{\xi}-\wt{\gm} & \doteq & \eta-\xi-\gm \end{array}\right\} \quad \Rightarrow\quad \nn \\ 
&& \left\{ \begin{array}{rcl} 
\xi(n) &=& \xi(0)+\left(\xi(0)-\eta(0)+\alpha+\Omega\right)n+\tfrac{1}{2}n(n-1)(\tau+\Omega') \\ 
\eta(n) &=& \eta(0)+\left(\xi(0)-\eta(0)+\alpha+\Omega\right)n+\tfrac{1}{2}n(n-3)(\tau+\Omega') \end{array}\right. 
\end{eqnarray}
\emph{Case iii)} 
\begin{eqnarray}\label{eq:N=1solsc}
&&\left.\begin{array}{rcl}
\wt{\xi}-\xi-\alpha & \doteq &\eta-\xi-\gm \\ 
\wt{\eta}-\wt{\xi}-\wt{\gm} & \doteq & \xi-\eta \end{array}\right\} \quad \Rightarrow\quad \nn \\ 
&& \left\{ \begin{array}{rcl} 
\xi(n) &=& \xi(0)+
\left(\alpha-\tfrac{1}{2}\gm_0-\tfrac{1}{4}\tau+\Omega'\right)n+\tfrac{1}{4}n(n-1)\tau  \\ 
 & &+\tfrac{1}{2}\left(\xi(0)-\eta(0)-\tfrac{1}{4}\tau+\tfrac{1}{2}(\gm_0+\Omega)\right)\left((-1)^n-1\right)  \\ 
\eta(n) &=& \eta(0)+
\left(\alpha-\tfrac{1}{2}\gm_0-\tfrac{1}{4}\tau+\Omega'\right)n+\tfrac{1}{4}n(n-3)\tau \\ 
 & &+\tfrac{1}{2}\left(\eta(0)-\xi(0)+\tfrac{1}{4}\tau-\tfrac{1}{2}(\gm_0+\Omega)\right)\left((-1)^n-1\right)  
\end{array}\right. 
\end{eqnarray}
\emph{Case iv)} 
\begin{eqnarray}\label{eq:N=1solsd}
&&\left.\begin{array}{rcl}
\wt{\xi}-\xi-\alpha & \doteq &\xi-\eta \\ 
\wt{\xi}-\wt{\eta} & \doteq & \eta-\xi-\gm \end{array}\right\} \quad \Rightarrow\quad \nn \\ 
&& \left\{ \begin{array}{rcl} 
\xi(n) &=& \xi(0)+
\left(\alpha-\tfrac{1}{2}(\gm_0+\Omega)-\tfrac{1}{4}\tau+\Omega'\right)n+\tfrac{1}{4}n(n-1)\tau  \\ 
 & &-\tfrac{1}{2}\left(\xi(0)-\eta(0)+\tfrac{1}{2}(\gm_0+\Omega)+\tfrac{1}{4}\tau\right)\left((-1)^n-1\right)  \\ 
\eta(n) &=& \eta(0)+
\left(\alpha-\tfrac{1}{2}(\gm_0+\Omega)-\tfrac{1}{4}\tau+\Omega'\right)n+\tfrac{1}{4}n(n-3)\tau \\ 
 & &+\tfrac{3}{2}\left(\eta(0)-\xi(0)-\tfrac{1}{4}\tau-\tfrac{1}{2}(\gm_0+\Omega)\right)\left((-1)^n-1\right)  
\end{array}\right. 
\end{eqnarray}
\ese 
in which $\Omega$ and $\Omega'$ denote arbitrary periods of the elliptic functions. Note that cases \emph{iii)} and \emph{iv)} differ from the 
cases \emph{i)} and \emph{ii)} by the appearance of alternating terms. As in the first order scheme we see that in the linearisable case 
there is quadratic dependence on the independent variable $n$, but in the higher order scheme it appears within a coupled system 
for $\xi(n)$ and $\eta(n)$.

\subsection{Case $N=2$} 

We will discuss the strategy to analyse the system of relations \eqref{eq:firsthc}, or, in the shorthand notation 
we introduced, comprising \eqref{rel:firstlast} together with \eqref{eq:matrixhigher} for the case 
$N=2$. As before, we assume the matrix $\bH$ to be of rank 1, where we can write as before $\bH=\bh^+(\bh^-)^T$. From 
\eqref{rel:firstlasta} we observe that either $\det(\bA_\tau)=0$ or that the matrix $[\bS]$ has to be singular (in fact, of rank 1 
when $N=2$). We pursue for convenience the latter case; in fact we will assume the coefficient $S_{ij}^{(l')}$ to be of the 
form  ~$S_{ij}^{(l')}=s^+_i s^0_{l'}s^-_j$~, in other words: fully factorized. Going back to the original form of the constitutive 
relations \eqref{eq:firsthc}, and inserting this \textit{Ansatz} into \eqref{eq:firsthca} we find that $s_j^-=\rho h_j^-$ for some 
factor $\rho$ and furthermore the following relation: 
\bse\label{eq:2ndordrels}\begin{equation} \label{eq:2ndordrelsa}
(H_1+H_2)\,\frac{\wt{S}^0}{S^0}\,\frac{\wt{S}_i}{H_i}=\sum_{l=1}^2\,(\bA_\tau)_{il}S_l\  ,   
\end{equation}
using as in section 2 the notation $S_l=s_l^-s_l^+$, $H_l=\wt{s}_l^-h_l^+$, and introducing $S^0:=s^0_1+s^0_2$. Eq. \eqref{eq:thirdhc} yields  
\begin{equation}\label{eq:2ndordrelsb}
\frac{\wt{S}_i}{H_i}\,\sum_{l=1}^2(\wt{\bK}_{-\tau})_{il}H_l(\bA_{-\tau})_{lj}=\sum_{l=1}^2 S_l(\bK_{-\tau})_{lj}\  ,  
\end{equation} 
in which   
\[ (\bK_{\kp})_{ij}:= \sum_{l'=1}^2 s_{l'}^0\,(\bE_{\kp})_{il'}(\bF_\kp)_{l'j}\  . \] 
From \eqref{eq:2ndordrelsb} it follows that the matrix $\bA_{-\tau}$ must be of rank 1, and hence we have again the condition 
\[ -\tau+\wt{\Xi}-\Xi-2\alpha\doteq 0\  .  \] 
Finally, from \eqref{eq:firsthcb} we have 
\begin{equation}\label{eq:2ndordrelsc}
\frac{\wt{S}_i}{H_i} \sum_{l=1}^2 H_l\,\left[(\wt{\bZ})_{il}+\wt{S}^0\zeta(\wt{\xi}_l-\xi_j-\alpha)\right]
=\sum_{l=1}^2 (\bA_\tau)_{il}S_l\,\left[S^0\zeta(\tau+\wt{\xi}_i-\xi_l-\alpha)+(\bZ)_{lj}\right]\  ,   
\end{equation} 
where the matrix $\bZ$ has entries 
\[ (\bZ)_{ij}:= \sum_{l'=1}^2 s_{l'}^0\,\left[ \zeta(\xi_i-\eta_{l'})+\zeta(\eta_{l'}-\xi_j-\gm)\right] \  . \]  
\ese 
The resolution of the system \label{eq:2ndordrels} can in principle be done following similar lines as the parallel system in 
section 3. However, in this case we have more variables at our disposal, including the coefficients $s^0_i$ and the intermediate 
variables $\eta_i$, ($i=1,2$), exploiting also the additional relation \eqref{eq:2ndordrelsb}. Elimination of the coefficient variables 
are expected to yield a coupled set of first order equations for $\xi_1(n)$ and $\eta_1(n)$, but the full analysis and assessment of the consequences, 
as well as generalizations to higher rank ($N\geq 3$) and higher order ($m\geq 3$), remain to be done and will be pursued in a follow-up 
paper.

\section{Conclusions}

In this paper we proposed a general system of elliptic discrete isomonodromic deformation problems from de-autonomisation of 
elliptic Lax pairs presented in our earlier paper \cite{nesliFrankRin}. While the first order scheme for $N=1$ and $N=2$ only leads 
to linearisable equations, we laid out the structure for the higher order (i.e., two-step) scheme and derived the constitutive relations 
by using some possibly novel elliptic identities. In principle the analysis in section 4 can be readily extended to the 3-step, or 
multi-step, case by using higher-order elliptic identities such as \eqref{eq:quadrupleprod} and the ones described in the Lemma of 
Appendix C. 
  
These constitutive relations contain coefficient matrices that are to be eliminated in order to yield a system of nonlinear 
non-autonomous elliptic ordinary difference equations which we expect to constitute higher-order and higher-rank 
versions of elliptic type Painlev\'e equations (i.e. elliptic Garnier and Schlesinger systems), but further analysis 
is needed to confirm those expectations. 
A general elliptic version of isomonodromic deformation theory was 
estblished some time ago by Krichever, \cite{Krich}, and a comparison with that work may establish that 
the systems proposed here are indeed isomonodromic in the sense of that theory. However, in that paper the 
compatibility conditions of the  isomonodromic system were not pursued, while here we have shown how to 
obtain a handle on that problem by using a system of elliptic identities which is excellently suitable for that purpose. The explicit formulae 
we obtained for the cases $N=1$ and $N=2$ confirm that the resulting equations have behaviour that one would expect from discrete 
Painlev\'e type equations, but further work is needed to make those assertions rigorous.    

We, furthermore, point out that, as a byproduct of the higher order isomonodromic scheme laid out in section 4, 
we can consider its autonomous limit, which amounts to a 2-step higher-time flow of the elliptic discrete-time Ruijsenaars model 
of \cite{NRK}. The Lax pair in that case is a discrete iso-spectral problem given by 
\be\label{eq:isospect} 
\bT'_\kp \bchi_\kp=\lambda \bchi_\kp\ , \quad \wt{\bchi}_\kp=\bL_\kp\,\bchi_\kp\  , \quad \wh{\bchi}_\kp=\bM_\kp\,\bchi_\kp\  , 
\ee
obtained by supplementing \eqref{eq:higherorderredcLax} with $\tau=0$ and $\gamma$ constant. In the stationary case the compatibility relations become
\begin{eqnarray} \label{highercasecompatibilityRS}
\wt{\bT}'_\kp\,\bL_\kp=\bL_{\kp}\,\bT'_\kp\ , \quad \wh{\bT}'_\kp\,\bM_\kp=\bM_{\kp}\,\bT'_\kp\ ,  
\end{eqnarray} 
together with  \eqref{eq:ZCcond}. 

The compatibility of \eqref{highercasecompatibilityRS} follows similar analysis as the one for non-autonomous case, making use of 
\eqref{eq:tripleprod} and the result is the following set of constitutive relations: 
\bse\label{higherconstitutiverelRS}\begin{eqnarray}
&&  \sum_{l,l'=1}^N H_{il}\,S_{lj}^{(l')}
=\sum_{l,l'=1}^N \wt{S}_{il}^{(l')}\,H_{lj}\  ,\label{higherconstitutiverelRSa}\\ 
&&\sum_{l,l'=1}^N H_{il}\,S_{lj}^{(l')}\,\frac{\sg(\wt{\xi}_i-\eta_{l'}-\alpha)\sg(\xi_l-\xi_j-\gamma)\sg(\wt{\xi}_i-\xi_l-\xi_j+\eta_{l'}-\alpha-\gamma)}{\sg(\wt{\xi}_i-\xi_l-\alpha)\sg(\xi_l-\eta_{l'})\sg(\eta_{l'}-\xi_j-\gamma)}\, \nn\\ 
&& \hspace{-.02cm} =\sum_{l,l'=1}^N \wt{S}_{il}^{(l')}\,H_{lj}\,
\frac{\sg(\wt{\xi}_i-\wt{\xi}_l-\gamma)\sg(\wt{\eta}_{l'}-\xi_j-\gamma-\alpha)\sg(\wt{\xi}_i+\wt{\xi}_l-\xi_j-\wt{\eta}_{l'}-\alpha)}{\sg(\wt{\xi}_i-\wt{\eta}_{l'})\sg(\wt{\eta}_{l'}-\wt{\xi}_l-\gamma)\sg(\wt{\xi}_l-\xi_j-\alpha)}\ ,\label{RSconst2eq}\nn \\
&& 
\end{eqnarray} 
and 
\begin{eqnarray} 
&& \sum_{l,l'=1}^N H_{il}\,S_{lj}^{(l')}\,\Phi_{\kappa_0}(\wt{\xi}_i-\xi_l-\alpha)\,\Phi_{\kappa_0}(\xi_l-\eta_{l'})\,\Phi_{\kappa_0}(\eta_{l'}-\xi_j-\gamma)\nn \\ 
&& \hspace{-.02cm} =\sum_{l,l'=1}^N \wt{S}_{il}^{(l')}\,H_{lj}\,\Phi_{\kappa_0}(\wt{\xi}_i-\wt{\eta}_{l'})\,\Phi_{\kappa_0}(\wt{\eta}_{l'}-\wt{\xi}_l-\gamma)
\,\Phi_{\kappa_0}(\wt{\xi}_l-\xi_j-\alpha)\ , 
\label{RSconst3eq} 
\end{eqnarray} \ese
where in the latter we can fix $\kappa_0$ to be any non-singular fixed value. The relations \eqref{higherconstitutiverelRSa} and \eqref{RSconst2eq} can be 
directly obtained by setting $\tau=0$ in the corresponding relations for the non-autonomous case, while \eqref{RSconst3eq} is just the Lax compatibility 
for any fixed value $\kp_0$ of the spectral parameter (in this case we cannot set $\kp_0=-\tau$ as in \eqref{eq:thirdhc}). We will leave the problem of 
attaining an explicit resolution of this system leading to closed-form expressions for higher discrete-time flows, to a future publication.

\subsection*{Appendix A: Weierstrass elliptic functions}
\def\theequation{A.\arabic{equation}}
\setcounter{equation}{0}

Here, we collect some useful formulae for elliptic functions, see also
the standard textbooks e.g. \cite{Akh,WW}.
The Weierstrass sigma-function is defined by
\be
\sigma(x) = x \prod_{(k,\ell) \ne (0,0)} (1-\frac{x}{\omega_{k\ell}})
 \exp\left[ \frac{x}{\omega_{k\ell}} + \frac{1}{2}
( \frac{x}{\omega_{k\ell}})^2\right]\ ,
\ee
with $\oa_{kl}=2k\oa_1 + 2\ell \oa_2$ and
$2\omega_{1,2}$  being a fixed pair of the primitive periods.
The relations between the Weierstrass elliptic functions are given by
\be
\zeta(x) =  \frac{\sigma^\prime(x)}{\sigma(x)}\   \  , \   \
\wp(x) = - \zeta^\prime(x)\   ,  \ee
where $\sg(x)$ and $\zeta(x)$ are odd functions and $\wp(x)$ is an
even function of its argument.
We recall also that the $\sg(x)$ is an entire function, and
$\zeta(x)$ is a meromorphic function having simple poles at
$\omega_{kl}$, both being quasi-periodic, obeying
\[ 
\zeta(x+2\omega_{1,2}) = \zeta(x) + 2\xi_{1,2}\    \ ,\    \
\sigma(x+2\omega_{1,2}) = -\sigma(x)
e^{2\xi_{1,2}(x+\omega_{1,2})}\  , 
\]
in which $\xi_{1,2}$ satisfy ~$\xi_1\omega_2 - \xi_2\omega_1
= \frac{\pi i}{2}$~, whereas $\wp(x)$ is doubly periodic.
The most important properties, for the sake of the computations in the main text,  
are the addition formulae, which are functional relations holding for arbitrary 
values (apart from singular points) for the variables in the arguments. 
The most fundamental is perhaps the three-term relation for $\sigma(x)$, which 
can be written as 
(\ref{eq:zs})
\bea
&& \sigma(x+a)\,\sigma(x-a)\,\sigma(y+b)\,\sigma(y-b)-\sigma(x+b)\,\sigma(x-b)\,\sigma(y+a)\,\sigma(y-a)  \nn  \\
&&\quad = \sigma(x+y)\,\sigma(x-y)\,\sigma(a+b)\,\sigma(a-b)\   .  \label{eq:8}
\eea
A limiting case of the latter is he relation 
\be \label{eq:zs}
\zeta(\ar) + \zeta(\bb) + \zeta(\gm) - \zeta(\ar +\bb + \gm)
  = \frac{  \sigma(\ar + \bb )  \sigma(\bb + \gm )
\sigma( \gm + \ar) }{ \sigma(\ar) \sigma(\bb) \sigma(\gm)
\sigma(\ar + \bb + \gm )}~ ,
\ee
between the $\sg$- and $\zeta$-functions. Furthermore, we have as a consequence of the latter
\be\label{eq:22}
\zeta(\ar+\bb) -\zeta(\ar)-\zeta(\bb)=\frac{1}{2}
\frac{\wp^\prime(\ar)-\wp^\prime(\bb)}{\wp(\ar)-\wp(\bb)}\   .
\ee
as well as the addition formula for the Weierstrass elliptic $\wp$-function: 
\be\label{eq:addform}
\Big(\zeta(\ar+\bb) -\zeta(\ar)-\zeta(\bb)\Big)^2= \wp(\ar)+\wp(\bb)+\wp(\ar+\bb)\  .
\ee
Finally we have the fundamental relation between $\sg$- and $\wp$-functions
\be\label{eq:sgP} 
\frac{\sg(x+y)\,\sg(x-y)}{\sg^(x)\,\sg^2(y)}=\wp(y)-\wp(x)\  , 
\ee 
which in turn gives back the three-term relation \eqref{eq:8} by using the identity
\[ \left(\wp(x)-\wp(a)\right)\left(\wp(y)-\wp(b)\right)-\left(\wp(x)-\wp(b)\right)\left(\wp(y)-\wp(a)\right)
=\left(\wp(x)-\wp(y)\right)\left(\wp(a)-\wp(b)\right)\  , \] 
thus, showing that no information gets lost if we reduce one functional relation to another.

\subsection*{Appendix B: The function $\Phi_\kp(x)$ and determinantal identities}
\def\theequation{B.\arabic{equation}}
\setcounter{equation}{0}

The function $\Phi(x)$ was introduced in \eqref{eq:Phi} and in terms of this function the various addition 
formulae of Appendix A can be conveniently expressed. Thus, eq. \eqref{eq:zs} can be cast into 
the form
\be\label{eq:12}
\Phi_\kp(x)\Phi_\kp(y) =
\Phi_{\kappa}(x+y)\left[ \zeta(\kp) +\zeta(x) +\zeta(y) -\zeta(\kp +x+y)
\right] \   ,
\ee
while \eqref{eq:8} can be rewritten as 
\be\label{eq:14}
\Phi_\kappa(x)\Phi_\ld(y) =
\Phi_{\kappa}(x-y)\Phi_{\kappa+\ld}(y) + \Phi_{\kappa +\ld}(x)
\Phi_{\ld}(y-x)\   ,
\ee
which can be considered as an elliptic analogue of the partial
fraction expansion. Furthermore, \eqref{eq:sgP} takes the form
\be\label{eq:16}
\Phi_\kp(x)\,\Phi_{-\kp}(x)=\wp(x)-\wp(\kp)\  . 
\ee  
Furthermore, in terms of this function we have the famous Frobenius formula, which can be 
considered to be an elliptic version of the well-known Cauchy determinantal identity. It reads:
\begin{equation}
\label{eq:cauchy}
\det\left( \Phi_\kappa(x_i - y_j)\right) =
\Phi_\kp( \Sigma ) \sg(\Sigma )
\frac{ \prod_{k<\ell} \sigma(x_k - x_\ell)
\sigma(y_\ell - y_k) }{
\prod_{k,\ell} \sigma(x_k - y_\ell) }\    \ ,\    \ {\rm where}\  \ \Sigma\equiv
\sum_i (x_i-y_i)\   ,
\end{equation}
cf. \cite{Frob}. From (\ref{eq:cauchy}), by expanding along one of its rows or columns,
an elliptic form of the Lagrange interpolation formula can obtained, which reads: 
\begin{equation}
\prod_{i=1}^N \frac{\sigma(\xi - x_i)}{\sigma(\xi - y_i)}\,=\,
\sum_{i=1}^N \Phi_{-\Sigma}(\xi - y_i)
\frac{\textstyle \prod_{j=1}^N \sigma(y_i - x_j)}{\textstyle
\prod_{j=1\atop j\ne i}^N \sigma(y_i - y_j)}\   ,
\label{eq:Lagr}
\end{equation}
for $\Sigma\neq 0$, where
\begin{equation}
\label{eq:Sigma}
\Sigma \equiv \sum_{i=1}^N (x_i -y_i)\   .
\end{equation}
When $\Sigma=0$ we recover the following formula
\begin{equation}
\prod_{i=1}^N \frac{\sigma(\xi - x_i)}{\sigma(\xi - y_i)}\,=\,
\sum_{i=1}^N \left[ \zeta(\xi - y_i)
- \zeta(x - y_i)\right]
\frac{\textstyle \prod_{j=1}^N \sigma(y_i - x_j)}{\textstyle
\prod_{j=1\atop j\ne i}^N \sigma(y_i - y_j)}\    ,
\label{eq:Lagr2}
\end{equation}
in which $x$ denotes any one of the zeroes $x_i$.
Note that in this case the left hand side is a meromorphic
function on the elliptic curve as a consequence of Abel's
theorem. Using (\ref{eq:Lagr}) it can be easily verified that eq.
(\ref{eq:Lagr2}) is independent of the choice of $x$.
In fact, this follows from the key property that
\begin{equation}
\sum_{i=1}^N
\frac{\textstyle \prod_{j=1}^N \sigma(y_i - x_j)}{\textstyle
\prod_{j=1\atop j\ne i}^N \sigma(y_i - y_j)}\,=\,0    ,
\label{eq:Lagr3}
\end{equation}
whenever ~$\sum_i(x_i-y_i)=0$~. This latter relation (\ref{eq:Lagr3}) is 
nothing else than a rewriting of (\ref{eq:Lagr}). Finally, we give the expression for the inverse of the elliptic
Cauchy matrix, namely
\begin{equation}
\label{eq:ACMinv}
\left[ \left( \Phi_\kp(x_\cdot - y_\cdot ) \right)^{-1}\right]_{ij} =
\Phi_{\kp+\Sigma}(y_i-x_j) \frac{ X(y_i) Y(x_j)}{Y_1(y_i)X_1(x_j)}\   ,
\end{equation}
(with $\Sigma$ as in \eqref{eq:Sigma}), in terms of the elliptic polynomials
$$ X(\xi) = \prod_{k=1}^N \sg(\xi -x_k) \     \ ,\     \
Y(\xi) = \prod_{k=1}^N \sg(\xi -y_k)\    , $$
and
\begin{equation}
X_1(x_j)=\prod_{k\neq j}\sigma(x_j-x_k)\,, \qquad
Y_1(y_i)=\prod_{k\neq i}\sigma(y_i-y_k)\,.
\end{equation}
Equation (\ref{eq:ACMinv}) can be derived using (\ref{eq:Lagr})
and (\ref{eq:Lagr2}).

\subsection*{Appendix C: Higher-order identities}
\def\theequation{C.\arabic{equation}}
\setcounter{equation}{0}

The addition formulae in terms of the function $\Phi_\kp(x)$ lend themselves fairly easily to a higher-order 
generalizations, which can be proven by induction from the basic ones. Thus, from \eqref{eq:14} one can prove the 
following general product identity 
\begin{equation}\label{eq:GenEllProd} 
\prod_{i=1}^n\,\Phi_{\kp_i}(x_i) =
\sum_{i=1}^n\,\Phi_{\kp_1+\cdots+\kp_n}(x_i)\,\prod_{j=1\atop j\neq i}^n\,\Phi_{\kp_j}(x_j-x_i)\  .
\end{equation}
Extending this identity to $n+1$ variables, including a $\kp_0$ and $x_0$, and subsequently taking the
limit $x_0=x_1+\varepsilon$, with $\varepsilon\to 0$, we obtain the following identity (after some obvious
relabelling of parameters and changes of variables):
\begin{eqnarray}
&& (-1)^{n-1}\Phi_{\kp_0+\kp_1+\cdots+\kp_n}(x_1+\cdots+x_n)\,\frac{\sg(x_1+\cdots+x_n)}{\prod_{j=1}^n\,\sg(x_j)} \nn \\
&& \times\left[ \zeta(\kp_0)+\sum_{j=1}^n\left(\zeta(\kp_j)+\zeta(x_j)\right)-
\zeta(\kp_0+\kp_1+\cdots+\kp_n+x_1+\cdots+x_n)\right] \nn \\
&=& \sum_{i=1}^n\,\Phi_{\kp_0+\kp_1+\cdots+\kp_n}(x_1+\cdots+\slashed{x_i}+\cdots+x_n)\,
\frac{\sg(x_1+\cdots+\slashed{x_i}+\cdots+x_n)\,\sg^{n-1}(x_i)}{\prod_{j=1\atop j\neq i}^n\,\sg(x_i-x_j)}\,
\prod_{j=0}^n\,\Phi_{\kp_j}(x_i)\  .  \nn \\ 
&& \label{eq:GenEllSum} 
\end{eqnarray}
These identities, which express sums of even numbers of $\zeta$-functions, are associated with the famous Frobenius-Stickelberger 
(i.e., an elliptic van der Monde) determinantal formula, \cite{FS}, which is given by: 
\begin{eqnarray}\label{eq:FrobStick}
&& \left| \begin{array}{cccccc}
1 & \wp(x_0) & \wp'(x_0) & \cdots & \cdots & \wp^{(n-1)}(x_0) \\
1 & \wp(x_1) & \wp'(x_1) & \cdots & \cdots & \wp^{(n-1)}(x_1) \\
\vdots & \vdots &  \vdots & \ddots & & \vdots \\
\vdots & \vdots &  \vdots & & \ddots & \vdots \\
1 & \wp(x_n) & \wp'(x_n) & \cdots & \cdots & \wp^{(n-1)}(x_n)
\end{array}\right| = \nn \\
&=& (-1)^{\frac{1}{2}n(n-1)} 1!2!\cdots n!\,
\frac{\sg(x_0+x_1+\cdots+x_n) \prod_{i<j=0}^n\sg(x_i-x_j)}{\sg^{n+1}(x_0)\,\sg^{n+1}(x_1)\cdots\sg^{n+1}(x_n)}\  .
\end{eqnarray}
A particular example of such an identity is the following one generalizing \eqref{eq:zs} to a 6-term relation: 
\begin{eqnarray}\label{eq:fourtermsigma} 
&& \quad\sigma(\kappa+x)\,\sigma(\lambda+x)\,\sigma(\mu+x)\sigma(\kappa+\lambda+\mu+y)\,\sigma^2(y) \nn \\
&& \quad\quad -\sigma(\kappa+y)\,\sigma(\lambda+y)\,\sigma(\mu+y)\sigma(\kappa+\lambda+\mu+x)\,\sigma^2(x) \nn \\
&& = \sigma(\kappa)\,\sigma(\lambda)\,\sigma(\mu)\,\sigma(x)\,\sigma(y)\,\sigma(\kappa+\lambda+\mu+x+y)\,\sigma(y-x) \nn \\
&& \quad\quad \times \left[\zeta(\kappa)+\zeta(\lambda)+\zeta(\mu)+\zeta(x)+\zeta(y)-\zeta(\kappa+\lambda+\mu+x+y) \right]
\end{eqnarray}
which derives from:
\begin{eqnarray}
&& \zeta(\kp)+\zeta(\ld)+\zeta(\mu)+\zeta(x)+\zeta(y)-\zeta(\kp+\ld+\mu+x+y)=  \nn \\
&& = \frac{\Phi_\kp(x)\Phi_\ld(x)\Phi_\mu(x)\Phi_{\kp+\ld+\mu}(y) - \Phi_\kp(y)
\Phi_\ld(y)\Phi_\mu(y) \Phi_{\kp+\ld+\mu}(x)}
{\Phi_{\kp+\ld+\mu}(x+y)\,(\wp(x)-\wp(y))} \  . \nn \\ 
&& \label{eq:6termsumzeta} 
\end{eqnarray}
In the same vein, we have the following 8-term $\zeta$-function relation:
\begin{eqnarray}\label{eq:4ordderrel}
&& \Phi_{\kp+\ld+\mu+\nu}(x+y+z)\,\frac{\sg(x+y+z)\,\sg(x-y)\,\sg(x-z)\,\sg(y-z)}{\sg^3(x)\,\sg^3(y)\,\sg^3(z)} \nn \\
&& \times \left[\zeta(\kp)+\zeta(\ld)+\zeta(\mu)+\zeta(\nu)+\zeta(x)+\zeta(y)+\zeta(z)-\zeta(\kp+\ld+\mu+\nu+x+y+z)\right]=  \nn \\
&=&  \Phi_\kp(x)\Phi_\ld(x)\Phi_\mu(x)\Phi_\nu(x)\left(\wp(z)-\wp(y)\right)\Phi_{\kp+\ld+\mu+\nu}(y+z) \nn \\
&&  + \Phi_\kp(y)\Phi_\ld(y)\Phi_\mu(y)\Phi_\nu(y)\left(\wp(x)-\wp(z)\right)\Phi_{\kp+\ld+\mu+\nu}(x+z) \nn \\
&&  + \Phi_\kp(z)\Phi_\ld(z)\Phi_\mu(z)\Phi_\nu(z)\left(\wp(y)-\wp(x)\right)\Phi_{\kp+\ld+\mu+\nu}(x+y)\  .
\end{eqnarray}
 
In the treatment of the main text of the paper we also need suitable formulae for multiple products of $\Phi_\kp(x)$-functions carrying the same 
index $\kp$. By expansion of the Frobenius formula \eqref{eq:cauchy} we find the following identity for products of $\Phi_\kp(x)$ functions 
with different arguments but with the same label $\kp$: 
\begin{equation}\label{eq:multiPhi}
\prod_{j=1}^n\,\Phi_\kp(x_j) = \tfrac{(-1)^{n-1}}{n-1}\Phi_\kp(x_1+\dots+x_n)\,
\frac{\left| \boldsymbol{1}\,,\,\wp(\boldsymbol{x})\,,\,\wp'(\boldsymbol{x})\,,\,\cdots\,,\,\wp^{(n-2)}(\boldsymbol{x})\right|}
{\left| \boldsymbol{1}\,,\,\tfrac{1}{2}\frac{\wp'(\boldsymbol{x})-\wp'(\kp)}{\wp(\boldsymbol{x})-\wp(\kp)}\,,\,\wp(\boldsymbol{x})\,,\,\wp'(\boldsymbol{x})\,,\,\cdots\,,\,\wp^{(n-3)}(\boldsymbol{x})\right|}\  ,
\end{equation}
where the r.h.s. contains a ratio of two $n\times n$ Frobenius-Stickelberger determinants and where each $f(\boldsymbol{x})$ stands for a function 
$f(x)$ denotes a column with entries $f(x_j)$ with $j=1,\dots,n$.

Particular examples of such identities, generalizing \eqref{eq:12} are the following higher-order relations:
\begin{eqnarray}\label{eq:tripleprod}
\Phi_\kp(x)\Phi_\kp(y)\Phi_\kp(z)&=& \tfrac{1}{2}\Phi_\kp(x+y+z)\,\left[ \left(\zeta(\kp)+\zeta(x)+\zeta(y)+\zeta(z)-\zeta(\kp+x+y+z)\right)^2 \right. \nn \\
                     && \left. \quad +\wp(\kp)-\left(\wp(x)+\wp(y)+\wp(z)+\wp(\kp+x+y+z) \right)\right] \  .
\end{eqnarray}
involving products of three $\Phi_\kp$ functions, and the next higher one reads: 
\begin{eqnarray}\label{eq:quadrupleprod} 
&& \Phi_\kp(x)\Phi_\kp(y)\Phi_\kp(z)\Phi_\kp(w)= \nn \\
&=&  \tfrac{1}{6}\Phi_\kp(x+y+z+w)\,\Big\{ \Big(\zeta(\kp)+\zeta(x)+\zeta(y)+\zeta(z)+\zeta(w)-\zeta(\kp+x+y+z+w)\Big)^3 \nn \\
&& -3\Big(\zeta(\kp)+\zeta(x)+\zeta(y)+\zeta(z)+\zeta(w)-\zeta(\kp+x+y+z+w)\Big) \nn \\
&& \qquad \times  \Big(\wp(x)+\wp(y)+\wp(z)+\wp(w)+\wp(\kp+x+y+z+w)-\wp(\kp)\Big)  \nn \\
&& -\Big(\wp'(\kp)+\wp'(x)+\wp'(y)+\wp'(z)+\wp'(w)-\wp'(\kp+x+y+z+w)\Big)  \Big\}\  .
\end{eqnarray}
The salient feature of the identities \eqref{eq:tripleprod} and \eqref{eq:quadrupleprod} is the way in which the label variable $\kp$ appears in the 
expressions between brackets on the right-hand sides: the $\kp$ appears on its own or in combination with the sums of all the arguments. 
The general structure of how these relations develop for higher and higher products is follows: 

\paragraph{\bf Lemma:} \emph{ 
The general form of identities of the type of products of the form 
\[ \prod_{j=1}^n \Phi_{\kp_j}(x) =:\tfrac{1}{(n-1)!} \mathcal{F}(\kp_1,\dots,\kp_n;x)\  , \]
is as follows. The function $\mathcal{F}$ is given by the expansion of the $(n-1)^{\rm th}$ derivative of the Weierstrass 
$\sg$-function divided by $\sg$ in terms of $\zeta$-functions and the $\wp$-function and its derivative, where whenever we 
have an odd function in this expansion (namely $\zeta$ and $\wp'$) we replace it by a combination of the form  
\[ \zeta(x)+\sum_{j=1}^n \zeta(\kp_j)-\zeta\left(\sum_{j=1}^n \kp_j+x\right)\  , \]
(and similar for $\wp'$) and when we encounter an even function ($\wp$ and powers of it) we replace it by 
\[ \sum_{j=1}^n \wp(\kp_j)+\wp\left(\sum_{j=1}^n \kp_j+x\right)-\wp(x)\  . \] 
} 

\paragraph{} In fact, if we inspect the first few cases of $\sg^{(n-1)}(x)/\sg(x)$: 
\[ \frac{\sg'(x)}{\sg(x)}=\zeta(x)\  , \quad   \frac{\sg''(x)}{\sg(x)}=\zeta^2(x)-\wp(x)\  ,\quad
\frac{\sg'''(x)}{\sg(x)}=\zeta^3(x)-3\zeta(x)\wp(x)-\wp'(x)\  , \dots  \]
expanded in $\zeta$, $\wp$ and $\wp'$, we see that they correspond exactly to the terms in the identities \eqref{eq:12}, 
\eqref{eq:tripleprod} and \eqref{eq:quadrupleprod}.


\begin{thebibliography}{99}
\bibitem {AC} 
M.J. Ablowitz and P.A. Clarkson,\emph{ Solitons, Nonlinear
  Evolution Equations and Inverse Scattering,} Cambridge Univ. Press, 1991.	

\bibitem {AdYam}
V.E.  Adler and R. Yamilov,\emph{ Explicit auto-transformations of integrable 
chains}, J. Phys. A \textbf{27} (1994), 477--492.

\bibitem {Adler2} V.E. Adler, \emph{B\"acklund transformation for the Krichever-Novikov equation}, 
Int. Math. Res. Not. \textbf{1} (1998), 1--4.

\bibitem {Adler3} 
V.E. Adler, \emph{ Discretizations of the Landau-Lifshits equation}, Theor. Math. Phys. 
\textbf{124} (1) (2000), 897--908. 

\bibitem{Akh} N.I. Akhiezer, \emph{Elements of the theory of elliptic functions}, Amer. Math. Soc., 1990.

\bibitem{ArinBor} 
D. Arinkin and A.Borodin, \emph{Moduli spaces of $d$-connections and difference Painlev\'e equations}, Duke Math. J. Vol. {\bf 134}, \# 3 (2006), 515-556.

\bibitem{Birk}
G.D. Birkhoff, \emph{General theory of linear difference equations}, Trans. Amer. Math. Soc. {\bf 12}, \# 2 (1911) 243--284. 

\bibitem{Birk2}
G.D. Birkhoff, \emph{The generalized Riemann problem for linear differential equations and the allied problem for linear difference and $q$-difference equations}, 
Proc. Amer. Acad. {\bf 49} (1913) 512--568. 

\bibitem{Bor} 
A. Borodin, \emph{Isomonodromy Transformations of Linear Systems of Difference Equations}, Ann. Math.(Series 2) Vol. {\bf 160} \# 3 (2004) 1141--1182.  

\bibitem{CJMNN} 
P. Clarkson, N. Joshi, M. Mazzocco, F.W. Nijhoff and M. Noumi, 
\emph{One hundred years of Painlev\'e VI, the Fuchs-Painlev\'e equation}, J.Phys. A:Math. Gen. {\bf 39} \# 39 (2006), special issue.   

\bibitem{nesliFrankRin} N. Delice, F.W. Nijhoff and S. Yoo-Kong, \emph{On elliptic 
Lax systems on the lattice and a compound theorem for hyperdeterminants}, J. Phys. A: Math. Theor. {\textbf {48(3)}} (2015), 035206.

\bibitem {FlasNew}H. Flaschka and A.C. Newell, \emph{Monodromy and spectrum-preserving deformations I.}, Comm. Math. Phys. \textbf{76} (1980), 65--116.
	
\bibitem{FoAb} A.S. Fokas and M.J. Ablowitz, \emph{Linearization of the Korteweg-de Vries and Painlev\'e II Equations,} Phys. Rev. Lett. \textbf{47} (1981), 1096--1100.

\bibitem{FIK} 
A.S. Fokas, A.R. Its and A. Kitaev, \emph{The isomonodromy approach to matrix models in 2D quantum gravity}, 
Commun. Math. Phys. {\bf 147} \# 2 (1992)  395--430. 

\bibitem{FIKN} 
A.S. Fokas, A.R. Its, A.A. Kapaev and V.Yu. Novokshenov, \emph{Painlev\'e Transcendents: The Riemann-Hilbert Approach}, (AMS Mathematical Surveys and Monographs, 2006).    
	
\bibitem{FS}
F.G. Frobenius and L. Stickelberger, \emph{Ueber die Addition und Multiplication der elliptischen Functionen}, J. Reine Angew. Math. {\bf 88} (1880), 146--184.

\bibitem{Frob} 
F.G. Frobenius,  \emph{Ueber die elliptischen Functionen zweiter Art}, J. Reine Angew. Math. {\textbf {93}} (1882), 53--68.

\bibitem{Fuch2}
R. Fuchs, \emph{Sur quelques \'equations diff\'erentielles lin\'eaires du second ordre}, C. R. Acad. Sci. (Paris) 141:555--558 (1905). 

\bibitem{Fuch}R. Fuchs, \emph{\"Uber lineare homogene Differentialgleichungen zweiter Ordnung mit drei im Endlichen gelegenen wesentlich singul\"aren Stellen}. Math. Ann. {\textbf {63}} (1907), 301--321.

\bibitem{Garnier}
R. Garnier, \emph{Sur des \'equations diff\'erentielles du troisi\`eme ordre dont l'int\'egrale g\'en\'erale est uniforme et sur une 
classe d'\'equations nouvelles d'ordre sup\'erieur}, Ann. \'Ecol. Norm. Sup., vol. 29:1--126 (1912). 


\bibitem{Carg} B. Grammaticos, F.W. nijhoff and A. Ramani, \emph{Discrete Painlev\'e Equations}, in: R. Conte (ed.), 
\emph{The Painlev\'e Property One Century Later}, CRM Series in Mathematical Physics, Springer-Verlag, (1999), pp. 413--516.

\bibitem{ItsNovok} 
A.R. Its, V.Yu. Novokshenov, \emph{The isomonodromic deformation method in the theory of Painlev\'e equations}, Lecture Notes in Mathematics 1191, (Berlin, New York, 1986). 


\bibitem{Jim1}M. Jimbo, T. Miwa and K. Uneo, \emph{Monodromy preserving deformations of linear ordinary differential equations with rational coefficients, I.}, 
 Physica, \textbf{2D} (1980) 306--352.

\bibitem{Jim2} M. Jimbo and T. Miwa, \emph{Monodromy preserving deformations of linear ordinary differential equations with rational coefficients, II.} Physica, \textbf{2D} (1981) 407--448.

\bibitem{Jim3} M. Jimbo and T. Miwa, \emph{Monodromy preserving deformations of linear ordinary differential equations with rational coefficients, III.} Physica, \textbf{4D} (1981) 26--46.

\bibitem{JimSakai}
M. Jimbo and H. Sakai, \emph{A $q$-analog of the sixth Painlev\'e equation}, Lett. Math. Phys. {\bf 38} \# 2 (1996) 145--154.  


\bibitem{JBH} 
N. Joshi, D. Burtonclay and R.G. Halburd, \emph{Nonlinear nonautonomous discrete dynamical systems from a general discrete isomonodromy problem}, 
Lett. Math. Phys., {\bf 26} (1992) 123-131.

\bibitem{KajNoumYam} K. Kajiwara, M. Noumi, and Y. Yamada.  \emph{Geometric Aspects of Painlev\'e Equations}, arXiv:1509.08186 (2015).

\bibitem{Krich}
I.M. Krichever, \emph{Analytic theory of difference equations with rational and elliptic coefficients and the Riemann-Hilbert problem}, Russ. Math. Surv. Vol. {\bf 59}, \# 6  
(2004).  

\bibitem{korsamt} D.A. Korotkin and J.A.H. Samtleben, \emph{On the quantization of isomonodromic deformations on the torus}, Intern. J. Mod. Phys.  \textbf{A12} (1997), 2013–2030.

\bibitem{Lax} 
P.D. Lax, \emph{Integrals of nonlinear equations of evolution and solitary waves}, Comm. Pure Applied Math. {\bf 21} (1968) 467--490. 

\bibitem{NijPap}
F.W. Nijhoff and V.G. Papageorgiou, \emph{Lattice Equations Associated with the Landau-Lifshitz Equations}, Physics Letter {\bf 141A} (1989) 269--274.

\bibitem{NP} 
F.W. Nijhoff and V.G. Papageorgiou, \emph{Similarity Reductions of Integrable Lattices and Discrete Analogues of the Painlev\'e II Equation}. Physics Letters {\textbf {A153}} (1991) 337--344.

\bibitem{NRK}
F.W. Nijhoff, O. Ragnisco and V. Kuznetsov, \emph{Integrable Time-Discretization
of the Ruijsenaars Model}, Comm. Math. Phys. {\textbf {176}} (1996) 681--700.

\bibitem{NRGO} F. W. Nijhoff, A. Ramani, B. Grammaticos and Y. Ohta,  \emph{On discrete Painlev\'e equations associated with the lattice KdV systems and the Painlev\'e VI equation,} Studies in applied mathematics {\textbf {106(3)}} (2001) 261--314.

\bibitem {NW} F.W. Nijhoff and  A.J. Walker, \emph{The discrete and continuous Painlev\'e  VI hierarchy and the Garnier systems,} Glasgow Math. J. {\bf{43A}} (2001) 109--123.

\bibitem{NoumTsujYam}  M. Noumi, S. Tsujimoto, and Y. Yamada, \emph{Pad\'e interpolation for elliptic Painlev\'e equation}, Symmetries, Integrable Systems and Representations, Springer Proceedings in Mathematics and Statistics {\textbf{40}} (2013) 463--482.

\bibitem{Okamoto2} K. Okamoto, \emph{On Fuchs's problem on a torus, I}, Funkcial. Ekvac. {\textbf{14}} (1971) 137--152.

\bibitem{Okamoto3} K. Okamoto, \emph{Sur le probl$\grave{e}$me de Fuchs sur un tore, II}, J. Fac. Sci. Univ. Tokyo, Sec. {\textbf{24(IA)}} (1977) 357--372.

\bibitem{Okamoto4} K. Okamoto, \emph{D\'eformation d'une \'equation diff\'erentielle lin\'eaire avec une singularit\'e irr\'eguli\`ere sur un tore}, J. Fac. Sci. Univ. Tokyo, Sec. {\textbf{26(IA)}} (1979)
 501--518.

\bibitem{OrmerodRains} 
C.M. Ormerod and E.M. Rains, \emph{Commutation relations and discrete Garnier systems}, preprint ArXiv:1601.06179. 

\bibitem{PNGR} V.G. Papageorgiou, F.W. Nijhoff, B.Grammaticos and A. Ramani, \emph{Isomonodromic deformation problems for discrete analogues of Painlev\'e equations}, 
Phys. Let. {\textbf{A164}} (1992) 57--64.

\bibitem{Rains}
E.M. Rains, \emph{An Isomonodromy Interpretation of the Hypergeometric Solution of the Elliptic Painlev\'e Equation (and Generalizations)}, SIGMA {\bf 7} (2011) 088--24 pages. 

\bibitem{Sakai} H. Sakai, \emph{Rational surfaces with affine root systems and geometry of the Painlev\'e equations}, Comm. Math. Phys. {\textbf{220}} (2001) 165--221.

\bibitem{Sakai2} H. Sakai, \emph{A $q$-analog of the Garnier system}, Funkcial. Ekvac. {\bf 48}, \# 2 (2005) 273--297.  

\bibitem{TongasNij2} 
A.S. Tongas and F.W. Nijhoff, \emph{A discrete Garnier type system from symmetry reduction on the lattice}, J.Phys.A:Math.Gen. {\bf 39} \# 39 
(2006) 12191--12202. 

\bibitem{Schlesin} L. Schlesinger, \emph{\"Uber eine Klasse von Differentialsystemen beliebiger Ordnung mit festen kritischen Punkten}, J. fur Math. (1912) {\textbf {141}} 96--145.

\bibitem{Takasaki}K. Takasaki \emph{Gaudin model, KZ Equation, and isomonodromic problem on torus}, Lett. Math. Phys. {\textbf{44(2)}} (1998) 143--156. 

\bibitem{Yamada} 
Y. Yamada, \emph{A Lax formalism for the elliptic difference Painlev\'e equation}, SIGMA {\textbf{5}} (2009), 042--15 pages. 

\bibitem{whittaker} E.T. Whittaker and G.N. Watson, \emph{ A course in modern analysis}, Cambridge University Press {\textbf{12}} (1988).

\end{thebibliography}
\end{document}